\documentclass[aps,pra,twocolumn,amsfonts,amssymb,amsmath,showpacs,floatfix,nofootinbib,citesort]{revtex4-2}
\usepackage{amsmath,amssymb,amsthm,mathtools}
\usepackage{mathrsfs}
\usepackage{bm}
\usepackage{bbm}
\usepackage{dsfont}

\usepackage{graphicx}
\usepackage{xcolor}
\usepackage{float}
\usepackage{textcomp}
\usepackage{CJK}
\usepackage[normalem]{ulem}

\definecolor{darkred}{rgb}{0.5,0,0}
\definecolor{darkblue}{rgb}{0,0,0.5}
\definecolor{darkgreen}{rgb}{0,0.5,0}

\usepackage{hyperref}
\hypersetup{
    pdftitle   = {Set Coherence},
    pdfauthor  = {},
    colorlinks = true,
    urlcolor   = blue,
    linkcolor  = red,
    citecolor  = blue,
    filecolor  = darkred
}

\theoremstyle{plain}
\newtheorem{theorem}{Theorem}
\newtheorem{lemma}{Lemma}
\newtheorem{pro}{Proposition}

\theoremstyle{definition}
\newtheorem{definition}{Definition}

\theoremstyle{remark}
\newtheorem{remark}{Remark}

\def\qed{%
    \leavevmode\unskip\penalty9999
    \hbox{}\nobreak\hfill
    \quad\hbox{\leavevmode
        \hbox to .77778em{%
            \hfil\vrule
            \vbox to .675em{%
                \hrule width .6em
                \vfil
                \hrule
            }%
            \vrule\hfil
        }%
    }%
    \par\vskip3pt
}

\newcommand{\be}{\begin{equation}}
    \newcommand{\ee}{\end{equation}}

\newcommand{\bex}{\begin{equation*}}
    \newcommand{\eex}{\end{equation*}}

\newcommand{\ba}{\begin{equation}\begin{aligned}}
        \newcommand{\ea}{\end{aligned}\end{equation}}

\newcommand{\bea}{\begin{eqnarray}}
    \newcommand{\eea}{\end{eqnarray}}

\newcommand{\beax}{\begin{eqnarray*}}
    \newcommand{\eeax}{\end{eqnarray*}}

\def\ra{\rangle}
\def\la{\langle}

\def\ot{\otimes}

\newcommand{\tr}{{\rm Tr}}
\newcommand{\diag}{\rm{diag}}

\newcommand{\mB}{\mathcal{B}}
\newcommand{\mC}{\mathcal{C}}
\newcommand{\mE}{\mathcal{E}}
\newcommand{\mF}{\mathcal{F}}
\newcommand{\mG}{\mathcal{G}}
\newcommand{\mH}{\mathcal{H}}
\newcommand{\mI}{\mathcal{I}}

\newcommand{\mP}{\mathcal{P}}

\newcommand{\mS}{\mathcal{S}}

\newcommand{\bu}{\mathbf{u}}
\newcommand{\bs}{\mathbf{s}}
\newcommand{\br}{\mathbf{r}}
\newcommand{\bn}{\mathbf{n}}

\newcommand{\bmm}{\mathbf{m}}

\newcommand{\epsbr}{\varepsilon_{\mathrm{br}}}

\newcommand{\mbR}{\mathbb{R}}

\begin{document}


\preprint{APS/123-QED}
\begin{CJK*}{GB}{gbsn}
\title{Measure of set coherence\\}


\author{Jiabo Pan}
\author{Yu Guo}
\email{guoyu3@aliyun.com}
\author{Yuqin Wang}

\affiliation{School of Mathematical Sciences, Inner Mongolia University, Hohhot, Inner Mongolia 010021, People's Republic of China}
\affiliation{Inner Mongolia Key Laboratory of Mathematical Modeling and Scientific Computing, Inner Mongolia University, Hohhot, Inner Mongolia 010021, People's Republic of China}


\begin{abstract}
Set coherence is a quantum resource for a set of states that characterizes whether a collection of quantum states contains coherence, namely whether these states fail to be simultaneously incoherent in a common basis. In this work, we show that the free operations of set coherence are common commutativity-preserving quantum channels, from which we then establish an axiomatic framework for set-coherence measure (SCM). Within this framework, we investigate the robustness based measures deduced by taking, respectively, the average, maximum, and sum of the single-state robustnesses over a common reference basis, where the first two were previously introduced in the literature and we check here whether they satisfy our SCM axioms. Consequently, we show that the averaged robustness is an SCM but it is neither unified nor complete, the maximal robustness is a unified SCM but is neither complete nor strongly superadditive, whereas the sum robustness is a complete SCM. Furthermore, we propose several classes of SCMs based on pairwise commutators. We show that the Schatten $p$-norm based SCMs are complete and strongly superadditive. We also find that the universal fourth-order Bargmann gap measure proposed in the literature is a complete and strongly superadditive quantifier of set coherence. Finally, we discuss the operational meaning of the SCM in the context of optimal approximate quantum broadcasting, and show that the amount of set coherence can closely reflect the performance of this task.


\end{abstract}

\maketitle
\end{CJK*}


\section{Introduction}


Quantum coherence is one of the most fundamental manifestations of quantumness~\cite{Streltsov2017}. Originating from the superposition principle, it underlies many nonclassical phenomena and serves as a useful resource in many quantum
information processing tasks that cannot be accomplished classically, such as quantum computing~\cite{Shor1997,Grover1997}, quantum cryptography~\cite{Grosshans2003}, quantum metrology~\cite{Giovannetti2011}, quantum thermodynamics~\cite{Lostaglio2015,Narasimhachar2015}, and quantum biology~\cite{Giovannetti2011}. In the standard resource theory of coherence, coherence is defined with respect to a fixed reference basis. The free states are density operators diagonal in this basis, while the free operations are those that cannot create coherence from incoherent states. This framework has led to a systematic understanding of quantum coherence~\cite{Baumgratz2014,Streltsov2017,Winter2016} and to many useful quantifiers, including the relative entropy of coherence~\cite{Baumgratz2014,Winter2016}, the robustness of coherence~\cite{Napoli2016}, observable coherence
measures~\cite{Girolami2014}, and asymmetry-based quantifiers~\cite{Marvian2016}, etc.

The basis-dependent formulation reveals a fundamental limitation: A single density operator does not possess coherence in an absolute sense, as it can always be diagonalized in its own eigenbasis. More critically, this single-state resource theory is inadequate for describing physical reality. In quantum physics, we must account for intrinsically collective phenomena---such as quantum pseudorandomness~\cite{Ji2018,Kretschmer2021,Bansal2025prx}, quantum broadcasting~\cite{Barnum1996,Barnum2007, Lu2013pra, Zhang2024, Buzek1997pra, Adhikari2008pra, Piani2008prl, Luo2010lmp, Luo2010pra, Mundra2019pra, Lostaglio2019prl, Marvian2019prl, Yang2022lma, Linan2024ctp, ApproxBroadcasting}, non-Markovian dynamics~\cite{Rivas2014rpp}, and linear-optical multiport interferometry~\cite{Shchesnovich2015pra,Shchesnovich2018pra,Jones2020,Jones2023pra}---none of which can be captured by isolated quantum states. This necessity compels us to consider structured families of quantum objects rather than individual instances~\cite{Designolle2021prl,Miyazaki2022q,Sajjan,Buscemi2020prl,Uola2019prl,Martins2020pra,Ducuara2020prr,Selby2023prl,Wagner2024pra,Galvao2020pra,Zhang2025arxiv,Zhang2025arxiv2,Gour2018,Guo2026pra}. Following this perspective, set coherence and set imaginarity emerged as key conceptual frameworks. Set coherence was rigorously formalized by Designolle \emph{et al.}~\cite{Designolle2021prl} building upon earlier works that explored analogous ideas under varied terminology~\cite{Horodecki2007pra,Galvao2020pra,Piani2014njp,Fuchs2003qic,Horodecki2006ijqi}. Independently, Miyazaki and Matsumoto ~\cite{Miyazaki2022q} introduced the notion of set imaginarity, later complemented by Guo \emph{et al.} in~\cite{Guo2026pra} wherein an axiomatic framework for quantifying it was proposed.

Set coherence has a clear operational meaning: a set of quantum states can be perfectly broadcast by a common quantum channel if and only if the states commute with each other~\cite{Barnum1996}. Recall that, for a set of states $\vec{\rho}=\{\rho_i\}_{i=1}^{n}$, the existence of a unitary $U$ such that all $U\rho_iU^\dagger$'s are simultaneously diagonal is equivalent to pairwise commutativity,
\[
[\rho_i,\rho_j]=0,
\qquad
\forall\,i,j.
\]
If this condition fails, the collection possesses a relational form of quantumness that cannot be attributed to any individual state. This basis-independent resource is referred to as set coherence~\cite{Designolle2021prl}. Namely, a set of states that is no-broadcasting is precisely the free-set condition of set coherence.

Set coherence therefore quantifies the obstruction to assigning a common classical description to a collection of quantum states. Recent invariant based approaches, particularly those based on Bargmann invariants, have provided experimentally accessible criteria
for detecting pairwise commutativity and hence set coherence~\cite{Wagner2026,Wang2026Bargmann}. Nevertheless, such criteria are
essentially binary: they determine whether a collection is free or resourceful, do not quantify the degree to which the states fail to admit a common incoherent basis, nor do they characterize how this resource transforms under physically admissible operations. A systematic resource theory of set coherence is therefore still needed.

In this work, we find that the free operations of set coherence coincide with a common class of commutativity-preserving quantum channels, which allows us to develop an axiomatic framework for set-coherence measures. We then examine the two robustness-based set-coherence measures proposed in~\cite{Designolle2021prl} against our axiomatic requirements. Noting that noncommutativity of the states is a signature of set coherence, we construct two classes of set-coherence measures based on the Schatten $p$-norm of pairwise commutators. We further assess, within our framework, the total Bargmann gap introduced in~\cite{Wang2026Bargmann}, which is derived from fourth-order Bargmann invariants. Finally, we derive an explicit quantitative relation between set coherence and the optimal approximate broadcasting error for pairs of pure qubit states.

Throughout this paper, we denote by $\mS(d)$ the set of all quantum states acting on the system described by the Hilbert space $\mH$ with $\dim\mH=d\geqslant2$. $\mP_{\mS(d)}$ denotes the set of all finite subsets in $\mS(d)$. For a set of states $\vec{\rho}$, we denote by $|\vec{\rho}|$ the cardinal number of $\vec{\rho}$. $\mB(\mH)$ denotes the Banach algebra of all linear operators acting on $\mH$ and $\mB_s(\mH)$ denotes the real space consisting of all self-adjoint operators in $\mB(\mH)$. For any $A\in\mB(\mH)$, $\|A\|_p$ denotes the Schatten $p$-norm of $A$, i.e., $\|A\|_p=\left( \tr|A|^p\right)^{1/p}$, where $|A|=(A^\dag A)^{1/2}$, $\|A\|=\sup_{|\psi\ra}\|A|\psi\ra\|=\lambda_{\max}$, where $\lambda_{\max}$ denotes the maximal eigenvalue of $|A|$, and $r(A)$ is the rank of $A$. $I^X$ (sometimes $I_X$ or $I$) denotes the identity operator on $\mH^X$.


\section{Preliminary}\label{sec2}


\subsection{Single state coherence}

In the resource theory of quantum coherence, the free state is identified with the incoherent state
\[
\rho=\sum_{i=1}^{d}\lambda_i |i\rangle\langle i|,
\]
i.e., a density operator whose matrix representation is diagonal in a fixed reference basis $\{|i\rangle\}_{i=1}^{d}$ of a $d$-dimensional Hilbert space $\mH$. The set of all incoherent states will be labeled by ${\mathcal I}$. The choice of this basis depends on the particular problem under study, and in many relevant scenarios such a basis is naturally singled out by the unavoidable decoherence~\cite{Zur}.

The corresponding free operations are called incoherent operations. A completely positive trace preserving map $\Phi$ is said to be an
incoherent operation if it admits a Kraus decomposition $\Phi(\rho)=\sum_j K_j\rho K_j^\dagger$, $\sum_j K_j^\dagger K_j=I$, such that, for every $\rho\in\mathcal I$ and every $j$, $\frac{K_j\rho K_j^\dagger}{\tr(K_j\rho K_j^\dagger)}\in\mathcal I$. Operationally, an incoherent operation can be interpreted as a quantum measurement whose individual outcomes cannot create coherence from an incoherent input, even when postselection on the measurement outcomes is allowed \cite{Baumgratz2014}. Kraus operators satisfying this requirement are called incoherent Kraus operators.

A nonnegative function $C$ on $\mS(d)$ is called a coherence measure if it satisfies the following conditions (C1) to (C4)~\cite{Streltsov2017}:

(C1) Non-negativity: $C (\rho) \geqslant 0$, and $C (\rho) = 0$ for any incoherent state $\rho$;

(C2) Monotonicity: $C(\mE(\rho))\leqslant C(\rho)$ whenever $\mE$ is incoherent operation;

(C3) Probabilistic monotonicity: $\sum_jp_jC(\rho_j)\leqslant C(\rho)$ for any set of incoherent operators $\{K_j\}$ satisfying $\sum_j K_j^{\dag}K_j=I$, where $p_j=\tr (K_j\rho K_j^{\dag})$, $\rho_j=\frac {1}{p_j}K_j\rho K_j^{\dag}$;

(C4) Convexity: $C(\sum_jp_j\rho_j)\leqslant \sum_jp_jC(\rho_j)$ for any set of states
$\{\rho_j\}$ and any $p_j\geqslant 0$ with $\sum_jp_j=1$.

\noindent We remark that (C3) together with (C4) implies (C2). There is also another framework for quantifying coherence~\cite{Yu2016pra}: Items (C1)-(C4) are equivalent to (C1), (C2) and

(C5) Additivity for direct sum states: $C(p\rho_1 \oplus (1-p)\rho_2) = pC(\rho_1)+(1-p)C(\rho_2)$,~$0<p<1$.

Several coherence measures have been proposed so far. We introduce here the distance based coherence measure and the robustness of coherence. Baumgratz \emph{et al}.~\cite{Baumgratz2014} introduced a general class of candidate distance based coherence measures,
\be
C_D(\rho)=\min_{\sigma\in\mI}D(\rho,\sigma),
\ee
where $D$ denotes a distance between quantum states. If $D$ is
contractive under completely positive trace preserving maps, then
$C_D$ is monotonic under incoherent operations. Important examples
include the relative entropy of coherence $C_r$~\cite{Baumgratz2014} and the $l_1$-norm of
coherence $C_{l_1}$~\cite{Baumgratz2014}, i.e.,
\bea
C_r(\rho)=\min\limits_{\delta\in\mI}S(\rho\|\delta)=S(\rho_{\diag})-S(\rho),\\
C_{l_1}(\rho)=\min\limits_{\delta\in\mI}\|\rho-\delta\|_{l_1}=\sum\limits_{i\neq j}|\la i|\rho|j\ra|,
\eea
where $S(\rho\|\delta)=\tr\rho(\log_2\rho-\log_2\delta)$ is the relative entropy, $\rho_{\diag}$ denotes the diagonal part of $\rho$.
In contrast, the candidate coherence quantifier induced by the squared Hilbert--Schmidt norm, i.e.,
\bea \label{C_l2}
C_{l_2}(\rho)=\min\limits_{\delta\in\mI}\|\rho-\delta\|_2^2=\sum\limits_{i\neq j}|\la i|\rho|j\ra|^2,
\eea
violates the monotonicity condition on average under selective incoherent operations~\cite{Baumgratz2014}. Rana \emph{et al}.~\cite{Rana2016pra} showed later that any coherence quantifier defined via the $l_p$ norm or the Schatten $p$-norm with $p \geqslant 2$ as in Eq.~\eqref{C_l2} violates (C3). For the case of $p=1$, i.e., the trace norm,
\beax
C_{tr}(\rho) = \min_{\delta \in \mathcal{I}} \|\rho - \delta\|_\tr
\eeax
satisfies the conditions (C1), (C2), and (C4), but the work of Yu \emph{et al}.~\cite{Yu2016pra} showed that the item (C5) may be violated.

For a single quantum state, robustness of coherence is typically defined as the minimum proportion of a noise state (or an arbitrary state) that must be mixed with the given state to transform it into a ``free state''~\cite{Napoli2016}, i.e.,
\be
C_R(\rho) = \min_{\tau} \left\{ s \geqslant 0 \,\bigg|\, \frac{\rho + s\tau}{1+s} \in \mI \right\},
\ee
where the minimization is taken over all quantum states $\tau$.

\subsection{Set coherence}

A set of quantum states
\be
\vec{\rho}=\{\rho_j:1\leqslant j\leqslant n\}\subseteq \mS(d)
\ee
is called set incoherent~\cite{Designolle2021prl} if there exists a unitary operator $U$ acting on $\mH$ such that all states $U\rho_jU^\dagger$ are diagonal with respect to some basis $\{|i\rangle\}_{i=1}^{d}$. Otherwise, the set $\vec{\rho}$ is said to have set coherence, or equivalently, $\vec{\rho}$ is called set coherent. Equivalently, $\vec{\rho}$ is set incoherent if and only if its elements commute pairwise, i.e.,
\beax
[\rho_i,\rho_j]=0,\quad\forall\, i,j.
\eeax

Designolle \emph{et al}.~\cite{Designolle2021prl} introduced two robustness based quantifiers of set coherence. One is the mean robustness of set coherence, which is defined as
\be
\mC'_R(\vec{\rho})=\min_U\frac{1}{n}\sum_{j=1}^{n}C_R(U\rho_jU^\dagger),
\ee
the other one is the max robustness of set coherence
\be
\mC''_R(\vec{\rho})=\min_U\max_{1\leqslant j\leqslant n}C_R(U\rho_jU^\dagger),
\ee
where $\mC_R$ denotes the robustness of coherence of a single quantum state with respect to the fixed reference basis.

\subsection{Broadcasting of quantum states}

Consider a bipartite system $AB$, where system $A$ is secretly prepared in one of the states $\rho_i$ from a set $\vec{\rho}=\{\rho_1, \rho_2, \dots, \rho_n\}$, while system $B$, which is intended to receive the unknown state, is initialized in a fixed standard state $\tau$. The initial state of the composite system is therefore $\rho_i\otimes\tau\in\mS^{AB}$, where $\mS^{AB}$ denotes the set of all states acting on $\mH^{AB}:=\mH^A\ot\mH^B$ with $\dim\mH^A=\dim\mH^B$. A physical process $\mathcal{E}$ transforms it as $\rho_i\otimes\tau \longmapsto \mathcal{E}(\rho_i\otimes\tau)=\varrho_i\in\mS^{AB}$.
The set $\vec{\rho}$ is said to be broadcast by $\mE$ if the output state $\varrho_i$'s satisfy~\cite{Barnum1996}
\[
\tr_A(\varrho_i)=\rho_i,\quad\tr_B(\varrho_i)=\rho_i,
\]
for all $i$'s. A special case of broadcasting is cloning~\cite{Scarani2005rmp}, for which the stronger condition $\mathcal{E}(\rho_i\otimes\tau)=\rho_i\otimes\rho_i$ is required. Thus, broadcasting only requires both reduced states to
coincide with the input state, while cloning requires the entire joint output state to be the product of two identical copies. It has been shown that there is no operation which can clone all quantum states~\cite{Wootters1982}.

The celebrated no-broadcasting theorem~\cite{Barnum1996} shows that a common physical process can broadcast $\vec{\rho}$ if and only if the states commute pairwise, namely,
\beax
\vec{\rho}\ \text{is perfectly broadcastable}\Longleftrightarrow [\rho_i,\rho_j]=0,\ \forall\,i,j.
\eeax
Namely, the set coherence coincides with such a no-go theorem: $\vec{\rho}$ can be perfectly broadcast if and only if it is set incoherent.

For pure states, the broadcasting condition becomes more restrictive. Let $\rho_i = |\psi_i\rangle\langle\psi_i|$ be a pure input state. If a bipartite output state $\varrho_i$ has a pure marginal, $\tr_B(\varrho_i)=|\psi_i\rangle\langle\psi_i|$, then the joint state cannot contain any correlation between $A$ and $B$, and must take the product form $\varrho_i=|\psi_i\rangle\langle\psi_i|\otimes\sigma_i^B$. If the second marginal is also required to reproduce the same input state, $\tr_A(\varrho_i)=|\psi_i\rangle\langle\psi_i|$, then necessarily $\sigma_i^B=|\psi_i\rangle\langle\psi_i|$, and therefore $\varrho_i=|\psi_i\rangle\langle\psi_i|\otimes|\psi_i\rangle\langle\psi_i|$. Hence, for pure states, perfect broadcasting is equivalent to perfect cloning. Moreover, for a pair of pure states, $\rho=|\psi\rangle\langle\psi|, \sigma=|\phi\rangle\langle\phi|$
the commutativity condition $[\rho,\sigma]=0$ holds if and only if the two states are either identical or orthogonal. Thus, a set of distinct pure states can be perfectly broadcast only if the states are mutually orthogonal.


\section{Free operation of set coherence}\label{sec3}


The conceptual framework of quantum resource theory rests on three fundamental pillars~\cite{Chitambar2019rmp,Gour2024arxiv,Coecke2016ic}: (i) the specification of a set of free states, where any state lying outside this set is treated as a resource state; (ii) a set of free operations, namely, operations that do not generate the resource in question; and (iii) a rigorous quantitative measure for assessing the resource. In the specific setting of set coherence, while the question of whether the set coherence is free or not has a clear answer, a systematic characterization of the corresponding free operations remains a notable gap in current research. A basic axiomatic requirement stipulates that any well-defined resource measure must be monotonically nonincreasing under admissible free operations. We thus discuss what is the free operation of set coherence first.

In Ref.~\cite{Guo2026pra}, we showed that the free operation of qubit set imaginarity is either any common unital operation or any common planarized but nonunital operation. Similarly, if a family of quantum operations $\{\mE_i\}$ acting on $\{\rho_i\}$, respectively, i.e., $\rho_i\mapsto\mE_i(\rho_i)$ for any $i$, these operations may create set coherence. For example, we consider a set incoherent pair $\vec{\rho}=(\rho,\sigma)$ with $\rho=\frac12(I+a\sigma_z)$ and $\sigma=\frac12(I+b\sigma_z)$, $0<|a|,|b|<1$. Clearly, $[\rho,\sigma]=0$. Let $U_1=I$ and $U_2(\theta)=e^{-\mathrm{i}\theta\sigma_y/2}$. Then
\[
U_2(\theta)\sigma U_2(\theta)^\dagger=\frac12\left[I+b\bigl(\sin\theta\,\sigma_x+\cos\theta\,\sigma_z\bigr)\right],
\]
and consequently
\[
\left[U_1\rho U_1^\dagger,U_2(\theta)\sigma U_2(\theta)^\dagger\right]=\frac{\mathrm{i}ab\sin\theta}{2}\sigma_y.
\]
Therefore, the transformed pair is not set incoherent whenever $\theta\notin\pi\mathbb Z$. So the free operation must be a common one acting on the set of states.

There is also an operational reason for adopting a common transformation. If a physical device receives an unknown state drawn from the set $\{\rho_i\}_{i=1}^n$ without access to the classical label $i$, then the device can implement only a single quantum channel $\mathcal{E}$, so that
\[
\rho_i \longmapsto \mathcal{E}(\rho_i),\qquad \forall\, i.
\]
By contrast, implementing a family of label-dependent channels $\{\mathcal{E}_i\}$ requires prior knowledge of the classical label $i$, and therefore corresponds to a different operational scenario. For this reason, we also need to restrict attention to common quantum channels that act componentwise on the entire state set.

For a quantum channel $\mathcal{E}$ and a set of states $\vec{\rho}=\{\rho_1,\ldots,\rho_n\}$, we write
\[
\mathcal{E}(\vec{\rho}):=\left\lbrace\mathcal{E}(\rho_1),\ldots,\mathcal{E}(\rho_n)\right\rbrace
\]
and denote by
\[
\mF_n(\mH):=\left\{\{\rho_1,\ldots,\rho_n\}:[\rho_i,\rho_j]=0\text{ for all }i,j, n\geqslant 2\right\}
\]
the collection of all set incoherent $n$ tuples of states. We now give the following definition.

\begin{definition}[Free operation of set coherence]
Let $\mathcal{E}:\mathcal{B}(\mathcal{H})\rightarrow\mathcal{B}(\mathcal{H})$ be a quantum channel. We call $\mathcal{E}$ a free operation of set coherence for $n$ tuples if $\vec{\rho}\in\mathcal{F}_n(\mathcal{H})$ implies $\mathcal{E}(\vec{\rho})\in\mathcal{F}_n(\mathcal{H})$, where $\mathcal{E}(\vec{\rho})=\{\mathcal{E}(\rho_1),\ldots,\mathcal{E}(\rho_n)\}$.
\end{definition}

As one may expect, the free operation of set coherence is precisely a commutativity-preserving channel.

\begin{theorem}\label{Characterization of free operations}
A quantum channel $\mathcal{E}:\mathcal{B}(\mathcal{H})\rightarrow\mathcal{B}(\mathcal{H})$ is a free operation of set coherence if and only if it is commutativity-preserving, i.e., $[\rho,\sigma]=0$ implies $[\mathcal{E}(\rho),\mathcal{E}(\sigma)]=0$ for any states $\rho$ and $\sigma$ acting on $\mH$.
\end{theorem}

\begin{proof}
Suppose first that $\mathcal{E}$ is commutativity-preserving. For any $\vec{\rho}=\{\rho_1,\ldots,\rho_n\}\in\mathcal{F}_n(\mathcal{H})$, we have $[\rho_i,\rho_j]=0$ for all $i,j$. Hence, $[\mathcal{E}(\rho_i),\mathcal{E}(\rho_j)]=0$ for all $i,j$, which implies $\mathcal{E}(\vec{\rho})\in\mathcal{F}_n(\mathcal{H})$.
Thus, $\mathcal{E}$ is a free operation of set coherence.

Conversely, if $\mathcal{E}$ is a free operation of set coherence, then for any states $\rho$ and $\sigma$ with $[\rho, \sigma]=0$, we have $[\mathcal{E}(\rho),\mathcal{E}(\sigma)]=0$, i.e., $\mathcal{E}$ is commutativity-preserving.
\end{proof}

We next briefly recall the concrete forms of commutativity-preserving quantum channels. We review some terminologies first. A channel $\mE:\mB(\mH)\rightarrow\mB(\mH)$ with $\dim\mH=d$ is called \emph{completely decohering}, or \emph{semiclassical}, if its range is commutative, namely,
$[\mE(A),\mE(B)]=0$ for all
$A,B\in\mB_s(\mH)$. $\mE$ is called \emph{isotropic} if
\be
\mE(A)
=
t\Gamma(A)
+
(1-t)\tr(A)\frac{I}{d},
\label{eq:isotropic channel}
\ee
where either $\Gamma(A)=UAU^\dagger$ or $\Gamma(A)=UA^{T}U^\dagger$ for some unitary operator $U$~\cite{Hu}. The parameter $t$ is restricted so that $\mE$ is completely positive and trace preserving. If $t\neq0$, the channel is called nontrivial. For $t=0$, $\mE$ reduces to the completely depolarizing channel, $\mE(A)=\tr(A)I/d$, and hence $\mE(\rho)=I/d$ for every quantum state $\rho$. The completely depolarizing channel is therefore a special case of a completely
decohering channel. For the case of $d\geqslant3$, a quantum channel is commutativity-preserving if and only if it is either completely decohering or a nontrivial isotropic channel~\cite{Guo2013jpa-qd}. Namely, whenever $d\geqslant3$, the free operations of set coherence are exactly these two classes of channels. The qubit case is different. For the qubit case, a quantum channel is commutativity-preserving if and only if it is either completely decohering or unital \cite{Guo2013jpa-qd}.

It is worth emphasizing that an incoherent operation defined with respect to a fixed reference basis need not be a free operation of set coherence, since the latter must preserve commutativity for arbitrary commuting input states. To illustrate this distinction, consider the qubit amplitude damping channel with Kraus operators
\[
K_0=
\begin{pmatrix}
    1&0\\
    0&\sqrt{1-\gamma}
\end{pmatrix},
~
K_1=
\begin{pmatrix}
    0&\sqrt{\gamma}\\
    0&0
\end{pmatrix},
~
0<\gamma<1.
\]
This channel is incoherent with respect to the computational basis. We take
$\rho_\pm=|\pm\rangle\langle\pm|$ with
$|\pm\rangle=(|0\rangle\pm|1\rangle)/\sqrt2$. Although
$[\rho_+,\rho_-]=0$, one obtains
\be
[\mE(\rho_+),\mE(\rho_-)]
=
\begin{pmatrix}
    0& -\gamma\sqrt{1-\gamma}\\
    \gamma\sqrt{1-\gamma}&0
\end{pmatrix}
\neq0.
\ee
That is, the amplitude damping channel can generate set coherence from a set incoherent pair and is not a free operation for set coherence.


\section{Axiomatic definition of set-coherence measure}\label{sec4}


Having characterized the free states and free operations of set coherence, we now introduce an axiomatic framework for quantifying it. Faithfulness and monotonicity under free operations are standard requirements for any resource measure. In addition, since enlarging a collection may introduce additional noncommuting pairs, it is natural to examine monotonicity under set inclusion. Intuitively, the union of two sets of states contains more set coherence than the sum of the two sets taken separately: if both $\vec{\rho}$ and $\vec{\sigma}$ are set incoherent and $\vec{\rho}\cap\vec{\sigma}=\emptyset$, but there exist some $\rho_i\in\vec{\rho}$ and $\sigma_j\in\vec{\sigma}$ with $[\rho_i, \sigma_j]\neq 0$, then $\vec{\rho}\cup\vec{\sigma}$ is not set incoherent. Moreover, if either $\vec{\rho}\cap\vec{\sigma}=\emptyset$ or $\vec{\rho}\cap\vec{\sigma}$ is additionally a commutative set, then $\vec{\rho}\cap\vec{\sigma}$ contains no set coherence, whereas the union $\vec{\rho}\cup\vec{\sigma}$ is set coherent. Together with the fact from Sec.~\ref{sec3} that the free operation of set coherence should be a common single operation, we establish the concept of set-coherence measure as follows.

\begin{definition}\label{SCM}
A function $\mC:\mP_{\mS(d)} \to\mbR_+$ is called a SCM if it
satisfies the following conditions (SC1)--(SC3):

(SC1) $\mC(\vec{\rho})=0$ iff $\vec{\rho}$ is set incoherent.

(SC2) For any unitary operator $U$,
\be
\mC(U\vec{\rho}U^\dagger)=\mC(\vec{\rho}),
\ee
where $U\vec{\rho}U^\dagger :=\{U\rho_jU^\dagger:1\leqslant j\leqslant n\}$.

(SC3) $\mC$ is monotonic under any commutativity-preserving operation $\mE$, i.e.,
\be
\mC\left[\mE(\vec{\rho})\right]\leqslant\mC(\vec{\rho})
\ee
for any $\vec{\rho}$.

Let $\mC$ be a SCM. If (SC4) for any $\vec{\rho}\subseteq\vec{\varrho}$,
\be
\mC(\vec{\rho})\leqslant\mC(\vec{\varrho}),
\ee
we call it a unified SCM. Moreover, for a unified SCM $\mC$ if (SC5) $\vec{\rho}\,\cap\,\vec{\sigma}=\emptyset$, then
\be
\mC(\vec{\rho})+\mC(\vec{\sigma})\leqslant \mC(\vec{\rho}\,\cup\,\vec{\sigma}),
\ee
we call it a complete SCM. Here we let $\mC(\emptyset)\equiv0$ and $\mC(\rho)\equiv0$ for any single state $\rho$. In addition, if a unified SCM $\mC$ satisfies (SC6)
\be
\mC(\vec{\rho})+\mC(\vec{\sigma})\leqslant \mC(\vec{\rho}\cup\vec{\sigma})+\mC(\vec{\rho}\cap\vec{\sigma}),
\ee
it is said to be strongly superadditive.
\end{definition}

Whenever $\mathcal{C}$ satisfies item (SC4), it is called set monotonic. Items (SC5) and (SC6) are termed superadditivity and strong superadditivity, respectively. Evidently, (SC6) implies (SC5), that is, strong superadditivity entails superadditivity.
These items and terminologies parallel those of the SIM introduced in~\cite{Guo2026pra}.

\begin{remark}
SCMs do not admit a counterpart of probabilistic monotonicity for the coherence measure. To see this, consider any free operation of set coherence with Kraus operators \(\{K_j\}\). Setting \(p_{ij}=\tr(K_j\rho_i K_j^\dagger)\), \(\vec{\rho}_{(j)}=\{\rho_{1j},\rho_{2j},\dots,\rho_{nj}\}\), and \(\rho_{ij}=\frac{1}{p_{ij}}K_j\rho_i K_j^\dagger\) for \(1\leqslant i\leqslant n\), we generally have \(p_{ij}\neq p_{i'j}\) whenever \(i\neq i'\). Only in the special case where, for each outcome \(j\), the probability \(p_{ij}\) is independent of the input state \(\rho_i\), i.e., \(p_{ij}=p_{i'j}\) for all \(1\leqslant i,i'\leqslant n\), may we write \(p_{ij}=p_j\) and impose the probabilistic monotonicity condition \(\sum_j p_j\mC(\vec{\rho}_{(j)})\leqslant\mC(\vec{\rho})\). The coherence measure is convex, but there is no counterpart of this convexity for SCMs, i.e. $\mC[\lambda\vec{\rho}+(1-\lambda)\vec{\sigma}]\leqslant\lambda\mC(\vec{\rho})+(1-\lambda)\mC(\vec{\sigma})$ is not generally valid for $|\vec{\rho}|=|\vec{\sigma}|$ and \(0<\lambda<1\). Indeed, let
$\vec{\rho}=\{\rho_z^+,I/2\}, \vec{\sigma}=\{I/2,\rho_x^+\},$
where $\rho_z^+=\frac12(I+\bm{\sigma_z})$, $\rho_x^+=\frac12(I+\bm{\sigma_x})$. Both $\vec{\rho}$ and $\vec{\sigma}$ are set incoherent, so the right-hand side of the inequality vanished (i.e., $\lambda\mC(\vec{\rho})+(1-\lambda)\mC(\vec{\sigma})=0$). However, for \(0<\lambda<1\), let $\omega_1=\lambda \rho_z^+ +(1-\lambda)\frac I2$,
$\omega_2=\lambda \frac I2+(1-\lambda)\rho_x^+$, then $\{\omega_1, \omega_2\} = \lambda\vec{\rho}+(1-\lambda)\vec{\sigma}$. We can easily check that
$[\omega_1,\omega_2]=\frac{{\rm i}\lambda(1-\lambda)}2 \bm{\sigma_y} \ne 0$.
So $\mC(\lambda\vec{\rho}+(1-\lambda)\vec{\sigma})\neq 0$. We thus conclude that any SCM cannot satisfy this general element-wise convexity.
\end{remark}

\begin{remark}
The terms unified SCM and complete SCM here are used in the same sense as unified SIM and complete SIM introduced in~\cite{Guo2026pra}, which are
in analogy with the unified multipartite entanglement measure (MEM) and complete MEM introduced in Refs.~\cite{Guo2020pra,Guo2022entropy,Guo2024pra,Guo2024rip,Guo2025arxiv} (see Remark 2 in ~\cite{Guo2026pra} for details).
\end{remark}


\section{Robustness of set coherence}\label{sec6}


We first investigate whether the previously proposed robustness-based quantities $\mC'_R$ and $\mC''_R$ in~\cite{Designolle2021prl} satisfy the SCM axioms established above.

\begin{theorem}\label{th2}
	$\mC'_R(\vec{\rho})$ is a SCM, but it is neither unified nor complete.
\end{theorem}

\begin{proof}
(SC1). Obviously, $\mC_R'(\vec{\rho})\geqslant0$. If $\vec{\rho}$ is set incoherent, then there exists a common unitary $U_0$ such that all $U_0\rho_jU_0^\dagger$ are incoherent in the fixed reference basis. Hence, $\mC_R(U_0\rho_jU_0^\dagger)=0$ for every $j$, and thus $\mC_R'(\vec{\rho})=0$. Conversely, we suppose that $\mC_R'(\vec{\rho})=0$. Since the unitary group is compact and the objective function is continuous, the minimum is attained by some unitary $U_0$. Therefore,
$\frac1n\sum_{j=1}^n
\mC_R(U_0\rho_jU_0^\dagger)=0$.
Since every term is nonnegative, $\mC_R(U_0\rho_jU_0^\dagger)=0$ for all $j$. Hence, all $U_0\rho_jU_0^\dagger$'s are incoherent in the same reference basis, so
$\vec{\rho}$ is set incoherent.

(SC2) is clear. In fact, for any unitary $V$, by setting $W=UV$, we obtain
$\mC_R'(V\vec{\rho}V^\dagger)
=
\min\limits_U\frac1n\sum_{j=1}^n
\mC_R(UV\rho_jV^\dagger U^\dagger)
=
\min\limits_W\frac1n\sum_{j=1}^n
\mC_R(W\rho_jW^\dagger)
=
\mC_R'(\vec{\rho})$.

(SC3). Let $\mE$ be a commutativity-preserving quantum channel. If $\mE$ is completely decohering, it is obvious that
\[
\mC_R'(\mE(\vec{\rho}))=0
\leqslant
\mC_R'(\vec{\rho}).
\]

Suppose now that $\mE$ is a nontrivial isotropic channel. We consider the case of $d\geqslant3$ first. Note that the robustness of coherence is monotonic under any completely positive and trace preserving (CPTP) map $\Phi$ satisfying
$\Phi(\mathcal I)\subseteq\mathcal I$, namely,
\bea\label{Cphi}
C_R(\Phi(\rho))\leqslant C_R(\rho), \quad \forall\,\rho\in\mS(d).
\eea
If $\mE(X)=tUXU^\dagger+(1-t)\tr(X)\frac{I}{d}$, $-\frac{1}{d^2-1}\leqslant t\leqslant1$,
where $U$ is a unitary operator, for any input unitary $W$, we choose
$V=WU^\dagger$. Then
\[
V\mE(\rho_j)V^\dagger
=
tW\rho_jW^\dagger
+
(1-t)\frac{I}{d}
=
\Phi_t(W\rho_jW^\dagger),
\]
where $\Phi_t(X)=tX+(1-t)\tr(X)\frac{I}{d}$. The channel $\Phi_t$ is CPTP in the above parameter range and preserves the set $\mathcal I$ of incoherent states. By Eq.~(\ref{Cphi}),
\bea\label{17}
C_R\!\left(V\mE(\rho_j)V^\dagger\right)
\leqslant
C_R(W\rho_jW^\dagger).
\eea

If $\mE(X)=tUX^{T}U^\dagger+(1-t)\tr(X)\frac{I}{d}$,
$-\frac{1}{d-1}\leqslant t\leqslant\frac{1}{d+1}$,
we choose $V=W^*U^\dagger$. Then
\[
V\mE(\rho_j)V^\dagger
=
t(W\rho_jW^\dagger)^T
+
(1-t)\frac{I}{d}
=
\Psi_t(W\rho_jW^\dagger),
\]
where $\Psi_t(X)=tX^T+(1-t)\tr(X)\frac{I}{d}$. In the stated parameter range, $\Psi_t$ is CPTP. Moreover, since $\delta^T=\delta$ for every $\delta\in\mathcal I$, one has $\Psi_t(\mathcal I)\subseteq\mathcal I$. Thus, Eq.~\eqref{17} also holds for this type of isotropic channels.

Consequently, for any type of isotropic channel,
\bea
\mC_R'(\mE(\vec{\rho}))
&=&
\min_U\frac1n\sum_{j=1}^n
C_R\!\left(U\mE(\rho_j)U^\dagger\right)
\nonumber\\
&\leqslant&
\frac1n\sum_{j=1}^n
C_R\!\left(V\mE(\rho_j)V^\dagger\right)
\nonumber\\
&\leqslant&
\frac1n\sum_{j=1}^n
C_R(W\rho_jW^\dagger).
\eea
Since this inequality holds for any unitary $W$, minimizing the right hand side over $W$ gives
\[
\mC_R'(\mE(\vec{\rho}))
\leqslant
\mC_R'(\vec{\rho}).
\]
Thus, $\mC_R'$ is monotonic under all nontrivial isotropic commutativity-preserving channels for the case of $d\geqslant3$.

For the qubit case, we only need to consider the case of unital qubit channels. For any qubit state $\rho$, we write it as
\[
\rho=\frac12\left(I+\br\cdot\boldsymbol{\sigma}\right),
\]
where $\br\in\mathbb R^3$ is its Bloch vector, $\boldsymbol{\sigma}=(\sigma_x, \sigma_y, \sigma_z)$ is the vector of Pauli operators. Let the reference basis be the eigenbasis of $\bn\cdot\boldsymbol{\sigma}$, where $\bn$ is a unit vector. Since the robustness of coherence coincides with the $l_1$-norm of coherence for qubit states~\cite{Napoli2016},
\be\label{robustness}
C_R^{(\bn)}(\rho)
=
\sqrt{|\br|^2-(\br\cdot\bn)^2}
=
\operatorname{dist}
\bigl(\br,\operatorname{span}\{\bn\}\bigr).
\ee
Hereafter, $C_R^{(\bn)}(\rho)$ denotes the robustness of coherence
with respect to the reference direction $\bn$ in the Bloch sphere. Thus, the robustness is precisely the length of the component of $\br$ perpendicular to the reference axis $\bn$.

Any unital qubit channel $\mE$ admits a Bloch representation $\br\mapsto T\br$, where $\|T\|\leqslant1$ is associated with $\mE$. We now fix an input reference direction $\bn$. If $T\bn\neq0$, we take $\bmm=\frac{T\bn}{|T\bn|}$. Then, for any qubit state $\rho$,
\bea
C_R^{(\bmm)}(\mE(\rho))
&=&
\operatorname{dist}
\bigl(T\br, \operatorname{span}\{T\bn\}\bigr)
\nonumber\\
&=&
\inf_{a\in\mathbb R}|T(\br-a\bn)|
\nonumber\\
&\leqslant&
\|T\|
\inf_{a\in\mathbb R}|\br-a\bn|
\nonumber\\
&\leqslant&
C_R^{(\bn)}(\rho).
\eea
If $T\bn=0$, we write
\[
\br=(\br\cdot\bn)\bn+\br_\perp,
\quad
\br_\perp\perp\bn.
\]
Then $T\br=T\br_\perp$. Choosing any unit vector $\bmm$, we
obtain
\bea
C_R^{(\bmm)}(\mE(\rho))
&=&
\operatorname{dist}
\bigl(T\br,\operatorname{span}\{\bmm\}\bigr)
\nonumber\\
&\leqslant&
|T\br|
=
|T\br_\perp|
\nonumber\\
&\leqslant&
\|T\||\br_\perp|
\leqslant
|\br_\perp|
\nonumber\\
&=&
C_R^{(\bn)}(\rho)
\eea
for any $\rho$.
\if false Thus, for any common input reference direction $\bn$, there exists a common output reference direction $\bmm$, independent of the state index $j$, such that
\[
\mC_R^{(\bmm)}(\mE(\rho_j))
\leqslant
\mC_R^{(\bn)}(\rho_j)
\]
for all $j$.

For qubit states, optimizing over common unitaries is equivalent to optimizing over the corresponding reference directions on the Bloch sphere. We therefore denote the optimization variable by $\bu$.
\fi
It turns out that
\bea
\mC_R'(\mE(\vec{\rho}))
&=&
\min_{\bu}\frac1n\sum_{j=1}^n
C_R^{(\bu)}(\mE(\rho_j))
\nonumber\\
&\leqslant&
\frac1n\sum_{j=1}^n
C_R^{(\bmm)}(\mE(\rho_j))
\nonumber\\
&\leqslant&
\frac1n\sum_{j=1}^n
C_R^{(\bn)}(\rho_j).
\eea
Minimizing the right hand side over $\bn$ gives $\mC_R'(\mE(\vec{\rho}))\leqslant\mC_R'(\vec{\rho}).$

(SC4). Let $\vec{\rho}=\{\rho_1,\ldots,\rho_n\}$ be any set-coherent collection that does not contain the maximally mixed state $\tau=I/d$ and
\[
\vec{\varrho}=\vec{\rho}\cup\{\tau\}.
\]
Since $U\tau U^\dagger=\tau$ and $C_R(\tau)=0$ for any unitary
$U$, we have
\beax
\mC_R'(\vec{\varrho})
&=&
\min_U\frac1{n+1}
\left[
\sum_{j=1}^n
C_R(U\rho_jU^\dagger)
+
C_R(U\tau U^\dagger)
\right]
\nonumber\\
&=&
\frac{n}{n+1}
\min_U\frac1n
\sum_{j=1}^n
C_R(U\rho_jU^\dagger)
\nonumber\\
&=&
\frac{n}{n+1}\mC_R'(\vec{\rho})
<
\mC_R'(\vec{\rho}),
\eeax
where the strict inequality follows from $\mC_R'(\vec{\rho})>0$. Although $\vec{\rho}\subseteq\vec{\varrho}$, one has $\mC_R'(\vec{\varrho})<\mC_R'(\vec{\rho})$. Hence, $\mC_R'$ is not monotonic under set inclusion and therefore is not a unified SCM.

(SC5). Let $\vec{\rho}=\{\rho_x^+, \rho_z^+\}$, $\vec{\sigma}=\{\tau_1, \tau_2\}$, where $\tau_1=\frac I2$, $\tau_2=\frac 12(I+\epsilon\bm{\sigma_x})$, $0<\epsilon<1$. Then $\vec{\rho}\cap\vec{\sigma}=\emptyset$ and $\vec{\rho}\cup\vec{\sigma}=\{\rho_x^+, \rho_z^+,\tau_1, \tau_2\}$. We can obtain $[\rho_x^+,\rho_z^+]=\frac{{\rm i}}{2}\sigma_{y}\neq 0$, $[\tau_1, \tau_2]=0$,
from Eq.~\eqref{robustness}, we know
\[
\mC'_R(\vec\rho)=\frac 12\min_{\bn}( \sqrt{1-n_x^2}+\sqrt{1-n_z^2} ).
\]
Taking $\bn=(1,0,0)$,  we can obtain
\beax
&&\mC_R'(\vec{\rho}\cup\vec{\sigma})\\
&\le& \frac{1}{4}\left[C_R^{(\bn)}(\rho_x^{+}) + C_R^{(\bn)}(\rho_z^{+}) + C_R^{(\bn)}(\tau_1) + C_R^{(\bn)}(\tau_2)\right]\\
&=&\frac 14,
\eeax
whereas
\[
\mC_R'(\vec{\rho})=\frac 12,  \quad \mC_R'(\vec{\sigma})=0.
\]
This leads to
$\mC_R'(\vec{\rho})+\mC_R'(\vec{\sigma})
>
\mC_R'(\vec{\rho}\cup\vec{\sigma})$,
that is, $\mC_R'$ is not a complete SCM.
\end{proof}

\begin{theorem}\label{pro6}
	$\mC_R''$ is a unified SCM, but it is neither complete nor strongly superadditive.
\end{theorem}

\begin{proof}
(SC1)-(SC2). If $\vec{\rho}$ is set incoherent, then there exists a unitary
$U_0$ such that all $U_0\rho_jU_0^\dagger$ are incoherent in some fixed
reference basis. Hence
$\max_jC_R(U_0\rho_jU_0^\dagger)=0$, and therefore
$\mC_R''(\vec{\rho})=0$. Conversely, suppose that
$\mC_R''(\vec{\rho})=0$. Since the unitary group is compact and the
objective function is continuous, the minimum is attained by some
unitary $U_0$. Thus
$\max_jC_R(U_0\rho_jU_0^\dagger)=0$. Since every term is
nonnegative, $C_R(U_0\rho_jU_0^\dagger)=0$ for all $j$, so all the
states are diagonal in one common basis.
(SC2) is clear.

(SC3). Let $\mE$ be a commutativity-preserving channel and $\vec{\rho}$ be a set of states. By the argument used in Theorem~\ref{th2}, for every unitary $U$, there exists a unitary $V=V(U)$, such that
$C_R(V\mE(\rho_j)V^\dagger)\leqslant
C_R(U\rho_jU^\dagger)$ for all $j$. Therefore,
\bea
\mC_R''(\mE(\vec{\rho}))
&=&
\min_W\max_jC_R(W\mE(\rho_j)W^\dagger)
\nonumber\\
&\leqslant&
\max_jC_R(V(U)\mE(\rho_j)V(U)^\dagger)
\nonumber\\
&\leqslant&
\max_jC_R(U\rho_jU^\dagger).
\eea
Since the above inequality holds for any $U$, minimizing the right-hand side over $U$ gives $\mC_R''(\mE(\vec{\rho}))\leqslant\mC_R''(\vec{\rho})$.

(SC4). Let $\vec{\rho}$ and $\vec{\varrho}$ be two sets of states with $\vec{\rho}\subseteq\vec{\varrho}$. Then for any unitary $U$, $\max_{\rho\in\vec{\rho}}C_R(U\rho U^\dagger)
\leqslant
\max_{\rho\in\vec{\varrho}}C_R(U\rho U^\dagger)$.
Taking the minimum over $U$ yields
$\mC_R''(\vec{\rho})\leqslant\mC_R''(\vec{\varrho})$. Thus,
$\mC_R''$ is a unified SCM.

(SC5)-(SC6). In order to show that $\mC_R''$ is not complete, we consider
$\vec{\rho}=\{\rho_x^+,\rho_z^+\}$ and
$\vec{\sigma}=\{\rho_x^-,\rho_z^-\}$, where
$\rho_x^\pm=\frac12(I\pm\sigma_x)$ and
$\rho_z^\pm=\frac12(I\pm\sigma_z)$. Clearly,
$\vec{\rho}\cap\vec{\sigma}=\emptyset$. Let the reference basis be determined by a unit Bloch vector $\bn=(n_x,n_y,n_z)$. Since     $C_R^{(\bn)}(\rho_x^\pm)=\sqrt{1-n_x^2}$ and $C_R^{(\bn)}(\rho_z^\pm)=\sqrt{1-n_z^2}$,
\[
\mC_R''(\vec{\rho})
=\min_{\bn} \max\{\sqrt{1-n_x^2},\sqrt{1-n_z^2}\}.
\]
Since $n_x^2+n_z^2\leqslant1$, at least one of $n_x^2$ and $n_z^2$ is not greater than $1/2$, and therefore the above maximum is at least $1/\sqrt2$. Equality is attained when $n_y=0$ and $n_x^2=n_z^2=1/2$. Thus,    $\mC_R''(\vec{\rho})=\mC_R''(\vec{\sigma})=1/\sqrt2$.

Moreover, $\vec{\rho}\cup\vec{\sigma}=\{\rho_x^+,\rho_x^-,\rho_z^+,\rho_z^-\}$. Since reversing a Bloch vector does not change its distance from the reference axis,     $\mC_R''(\vec{\rho}\cup\vec{\sigma})=1/\sqrt2$. This leads to
\[
\mC_R''(\vec{\rho})+\mC_R''(\vec{\sigma})
=\sqrt2
>\frac1{\sqrt2}
=\mC_R''(\vec{\rho}\cup\vec{\sigma}),
\]
which violates the superadditivity condition for disjoint sets. Hence,
$\mC_R''$ is not a complete SCM. With the convention $\mC_R''(\emptyset)=0$, the same example also violates
\[
\mC_R''(\vec{\rho})+\mC_R''(\vec{\sigma})
\leqslant
\mC_R''(\vec{\rho}\cup\vec{\sigma})
+ \mC_R''(\vec{\rho}\cap\vec{\sigma}),
\]
so $\mC_R''$ is not strongly superadditive.
\end{proof}

By the arguments in Theorem~\ref{th2}, the multiplier factor ``$1/n$'' in the definition of $\mC'_{R}$ leads to the violation of items (SC4)--(SC6). We thus define
\[
\mC'''_R(\vec{\rho})
:=
\min_U\sum_{j=1}^{n}
\mC_R(U\rho_jU^\dagger)
\]
with respect to some given reference basis.

\begin{theorem}\label{pro4}
	$\mC'''_R$ is a complete SCM, but it is not strongly
	superadditive.
\end{theorem}

\begin{proof}
Items (SC1)-(SC3) are clear following the argument in Theorem~\ref{th2}.

(SC4). If $\vec{\rho}\subseteq\vec{\varrho}$, then for any unitary $U$,
\[
\sum_{\rho_i\in\vec{\rho}}C_R(U\rho_iU^\dagger)
\leqslant
\sum_{\varrho_j\in\vec{\varrho}}C_R(U\varrho_jU^\dagger),
\]
and hence
$\mC'''_R(\vec{\rho})
\leqslant\mC'''_R(\vec{\varrho})$.

(SC5). For disjoint sets $\vec{\rho}$ and $\vec{\sigma}$,
\bea
\mC'''_R(\vec{\rho}\cup\vec{\sigma})
&=&
\min_U\left[
\sum_{\rho_i\in\vec{\rho}}\mC_R(U\rho_iU^\dagger)
+
\sum_{\sigma_j\in\vec{\sigma}}\mC_R(U\sigma_jU^\dagger)
\right]
\nonumber\\
&\geqslant&
\mC'''_R(\vec{\rho})
+
\mC'''_R(\vec{\sigma}).
\eea
Therefore, $\mC'''_R$ is a complete SCM.

(SC6) To show that it is not strongly superadditive, we consider
$\vec{\rho}=\{\rho_z^+,\rho_x^+\}$ and
$\vec{\sigma}=\{\rho_z^+,\rho_x^-\}$. For a reference direction
$\bn=(n_x,n_y,n_z)$,
\[
C_R^{(\bn)}(\rho_z^+)=\sqrt{1-n_z^2},
\quad
C_R^{(\bn)}(\rho_x^\pm)=\sqrt{1-n_x^2}.
\]
Hence,
\[
\mC'''_R(\vec{\rho})
=
\mC'''_R(\vec{\sigma})
=
\min_{\bn}
\left[
\sqrt{1-n_z^2}+\sqrt{1-n_x^2}
\right]
=1.
\]

Moreover, $\vec{\rho}\cap\vec{\sigma}=\{\rho_z^+\}$, so
$\mC'''_R(\vec{\rho}\cap\vec{\sigma})=0$, while
\[
\mC'''_R(\vec{\rho}\cup\vec{\sigma})
=
\min_{\bn}
\left[
\sqrt{1-n_z^2}
+
2\sqrt{1-n_x^2}
\right]
=1.
\]
To see the latter, note that the minimum occurs for $n_y=0$ and $n_x^2+n_z^2=1$. Writing $n_x=\cos\theta$ and $n_z=\sin\theta$ gives $\cos\theta+2\sin\theta$, whose minimum on $[0,\pi/2]$ is $1$, attained at $\theta=0$. Consequently,
\[
\mC'''_R(\vec{\rho})
+
\mC'''_R(\vec{\sigma})
= 2
>1
=\mC'''_R(\vec{\rho}\cup\vec{\sigma})
+\mC'''_R(\vec{\rho}\cap\vec{\sigma}),
\]
which violates strong superadditivity.
\end{proof}

Comparing with the robustness-based SIMs $\check{\mI}_R$, $\check{\mI}'_R$ and $\check{\mI}''_R$ proposed in~\cite{Guo2026pra}, we find that $\mC'_R$, $\mC''_R$, and $\mC'''_R$ are the counterparts of $\check{\mI}_R$, $\check{\mI}'_R$, and $\check{\mI}''_R$, respectively. However, none of these measures is strongly superadditive even though some of them--- $\check{\mI}''_R$ and $\mC'''_R$---are complete measures. In the next section, we will propose two classes of SCMs which are not only complete but also strongly superadditive.


\section{Set coherence measures based on commutator norms}\label{sec5}


Since a collection of quantum states is set incoherent if and only if
all of its elements commute pairwise \cite{Designolle2021prl,Wagner2026}, the commutators $[\rho_i,\rho_j]$ provide a natural algebraic witness of set coherence. Therefore, a nonzero commutator directly quantifies the obstruction to such a common
commutative description. This motivates the use of norms of $[\rho_i,\rho_j]$ to construct quantitative measures of set coherence.

For any set of quantum states $\vec{\rho}=\{\rho_j:1\leqslant j\leqslant n\}$, we define
\be\label{eq:C1}
\mC_p(\vec{\rho})
:=\sum_{i<j}\|[\rho_i,\rho_j]\|_p,\quad 1\leqslant p<\infty.
\ee
We next show that $\mC_p$ is a well-defined SCM for any $1\leqslant p<\infty$. We first give the following Lemma, whose proof is given in Appendix~\ref{a}, which is necessary in our discussion.

\begin{lemma}\label{monotonicity}
	Let $\mE$ be a commutativity-preserving quantum channel acting on a
	finite dimensional quantum system. Then, for any quantum states
	$\rho$ and $\sigma$ and any $1\leqslant p\leqslant\infty$,
	\be
	\bigl\|[\mE(\rho),\mE(\sigma)]\bigr\|_p
	\leqslant
	\bigl\|[\rho,\sigma]\bigr\|_p.
	\ee
\end{lemma}

\begin{theorem}\label{thm:C1C2}
    $\mC_p$ is a complete SCM and is strongly superadditive, $1\leqslant p<\infty$.
\end{theorem}

\begin{proof}
Since every term is nonnegative and a norm vanishes only at the zero operator, $\mC_p(\vec{\rho})=0$ if and only if $[\rho_i,\rho_j]=0$ for all $i,j$. Thus, $\mC_p(\vec{\rho})=0$ if and only if $\vec{\rho}$ is set incoherent. For any unitary operator $U$, the unitary invariance of the Schatten $p$-norms gives $\mC_p(U\vec{\rho}U^\dagger)=\sum_{i<j}
\|[U\rho_iU^\dagger,U\rho_jU^\dagger]\|_p=\sum_{i<j}\|U[\rho_i,\rho_j]U^\dagger\|_p =\mC_p(\vec{\rho})$. Let $\mE$ be a commutativity-preserving operation. By Lemma~\ref{monotonicity}, $\mC_p\bigl(\mE(\vec{\rho})\bigr)=\sum_{i<j}
\|[\mE(\rho_i),\mE(\rho_j)]\|_p \leqslant\sum_{i<j}\|[\rho_i,\rho_j]\|_p =\mC_p(\vec{\rho})$. That is, $\mC_p$ is monotonic under any free operation of set coherence.

If $\vec{\rho}\subseteq\vec{\varrho}$, every pair appearing in $\mC_p(\vec{\rho})$ also appears in $\mC_p(\vec{\varrho})$, while all additional terms are nonnegative. Therefore, $\mC_p(\vec{\rho})\leqslant\mC_p(\vec{\varrho})$, and hence $\mC_p$ is a unified SCM.

For two sets of states $\vec{\rho}$ and $\vec{\sigma}$, we let $A=\vec{\rho}\setminus\vec{\sigma}$ and $B=\vec{\sigma}\setminus\vec{\rho}$. By decomposing the corresponding pairwise sums, we obtain
\beax
&&
\mC_p(\vec{\rho}\cup\vec{\sigma})
+
\mC_p(\vec{\rho}\cap\vec{\sigma})
-
\mC_p(\vec{\rho})
-
\mC_p(\vec{\sigma})
\nonumber\\
&=&
\sum_{\rho_i\in A,\ \sigma_j\in B}
\|[\rho_i,\sigma_j]\|_p
\geqslant0.
\eeax
Thus,
$\mC_p(\vec{\rho})+\mC_p(\vec{\sigma})
\leqslant
\mC_p(\vec{\rho}\cup\vec{\sigma})
+
\mC_p(\vec{\rho}\cap\vec{\sigma})$,
so $\mC_p$ is strongly superadditive. In particular, if $\vec{\rho}\cap\vec{\sigma}=\emptyset$, then $\mC_p(\vec{\rho})+\mC_p(\vec{\sigma})\leqslant\mC_p(\vec{\rho}\cup\vec{\sigma})$, which proves that $\mC_p$ is a complete SCM.
\end{proof}

Recently, Wang~\cite{Wang2026Bargmann} established a universal fourth order Bargmann invariant criterion for pairwise commutativity: For any two states $\rho$ and $\sigma$, $\Gamma(\rho, \sigma):=\tr(\rho^2\sigma^2)-\tr(\rho\sigma\rho\sigma)=\frac12\|[\rho,\sigma]\|_2^2$, namely $\Gamma(\rho,\sigma)=0$ if and only if $[\rho, \sigma]=0$. For any set of quantum states $\vec\rho=\{\rho_j:1\leqslant j\leqslant n\} \subseteq \mS(d)$, they defined the total ``Bargmann Gap'' as~\cite{Wang2026Bargmann}
\[
\mG_B(\vec\rho)=\sum_{i<j}\Gamma(\rho_i,\rho_j).
\]
By definition, it is indeed a variation of $\mC_p$ with $p=2$. By the relation $\Gamma(\rho, \sigma)=\frac12\|[\rho,\sigma]\|_2^2$, following the same argument as in Theorem \ref{thm:C1C2}, we can conclude the following.

\begin{pro}
$\mG_B$ is a complete SCM and is strongly superadditive.
\end{pro}

Similarly, we can define
\be\label{eq:Cp'}
\mC_p'(\vec{\rho})
:=\sum_{i<j}\|[\rho_i,\rho_j]\|_p^p,\quad 1\leqslant p<\infty.
\ee
It is also a complete SCM and is strongly superadditive. In
contrast to coherence measures, for which only the $l_1$-norm yields a valid measure, any Schatten $p$-norm can be used to
quantify set coherence. We next estimate the $p$-norm of the commutator $[\rho, \sigma]$ for any pair of $\rho$ and $\sigma$, from which we obtain an upper bound for these SCMs for any given set of states.

\begin{pro}
For any two quantum states $\rho$ and $\sigma$ and
$1\leqslant p<\infty$,
\[
\|[\rho,\sigma]\|_p\leqslant 2^{1/p-1}.
\]
In particular, the bound is attained by two pure states $\rho=|\psi\ra\la\psi|$ and $\sigma=|\phi\ra\la\phi|$ satisfying
$|\langle\psi|\phi\rangle|^2=1/2$, and
\[
\mG_B(\rho,\sigma)\leqslant\frac14.
\]
\end{pro}

\begin{proof}
For fixed $\sigma$, the map $\rho\mapsto\|[\rho,\sigma]\|_p$ is convex. Hence its maximum over $\rho$ can be attained at a pure state. Fixing such a pure state and
applying the same argument to $\sigma$, we conclude that the joint maximum can be attained at a pair of pure states.

Let $\rho=|\psi\rangle\langle\psi|$ and $\sigma=|\phi\rangle\langle\phi|$, and set
$c=|\langle\psi|\phi\rangle|^2$. Up to some phase, we write $|\phi\rangle=\sqrt{c}\,|\psi\rangle+\sqrt{1-c}\,|\psi^\perp\rangle$, where $|\psi^\perp\ra$ is a vector in the orthocomplement space of $|\psi\ra$. On
$\operatorname{span}\{|\psi\rangle,|\psi^\perp\rangle\}$,
\[
[\rho,\sigma]
=
\sqrt{c(1-c)}
\begin{pmatrix}
	0&1\\
	-1&0
\end{pmatrix}.
\]
Thus, its two nonzero singular values are both $\sqrt{c(1-c)}$, and therefore
\[
\|[\rho,\sigma]\|_p
=
2^{1/p}\sqrt{c(1-c)}
\leqslant
2^{1/p-1}.
\]
The upper bound is attained when $c=1/2$. Finally, $\Gamma(\rho,\sigma)=\frac12\|[\rho,\sigma]\|_2^2$ gives $\Gamma(\rho,\sigma)\leqslant1/4$.
\end{proof}

\begin{remark}
	For any $\vec{\rho}\subset\mS(d)$ and $\vec{\varrho}\subset\mS(d')$ with $|\vec{\rho}|=|\vec{\varrho}|$, $0<\lambda<1$, we let
	\beax
	\lambda\vec{\rho} \oplus (1-\lambda)\vec{\varrho}&:=&\{\lambda\rho_1\oplus(1-\lambda)\varrho_1, \lambda\rho_2\oplus(1-\lambda)\varrho_2,\\
	&&~\dots, \lambda\rho_n\oplus(1-\lambda)\varrho_n\}.
	\eeax
	We write $\omega_i = \lambda\rho_i \oplus (1-\lambda)\varrho_i$, then
	\[
	\begin{aligned}
		[\omega_i,\omega_j]
		&= \big[\lambda\rho_i \oplus (1-\lambda)\varrho_i,\; \lambda\rho_j \oplus (1-\lambda)\varrho_j\big] \\
		&= \lambda^2[\rho_i,\rho_j] \oplus (1-\lambda)^2[\varrho_i,\varrho_j],
	\end{aligned}
	\]
	which reveals that $\vec{\omega}=\{\omega_i\}$ is set incoherent if and only if both $\vec{\rho}$ and $\vec{\varrho}$ are set incoherent. But for any SCM $\mC$, $\mC(\lambda\vec{\rho} \oplus (1-\lambda)\vec{\varrho})\neq \lambda \,\mC(\vec{\rho}) + (1-\lambda)\mC(\vec{\varrho})$ in general. For example, the inequality fails already for $\mC=\mC_p$ or $\mG_B$. Namely, there is no counterpart of direct sum additivity for SCM.
\end{remark}

We note a simple property: every unitary operation is necessarily a free operation of set coherence, and it freezes all set-coherence measures. We now go one step further and address the converse question: must a free operation of set coherence that freezes $\mC_p$ be unitary?

\begin{pro}\label{pro3}
Let $\mE$ be a commutativity-preserving quantum channel on $\mS(d)$ with $\dim\mH=d\ge2$. Then
\bea\label{freezing}
\mC_p(\mE(\vec{\rho}))=\mC_p(\vec{\rho}),\quad \forall\,\vec{\rho}
\eea
if and only if $\mE$ is a unitary channel, namely,
\[
\mE(X)=UXU^\dagger
\]
for some unitary operator $U$. The same conclusion holds for $\mC_p'$.
\end{pro}

\begin{proof}
We only need to prove the necessity.

\textit{Case 1: $d=2$.} Every commutativity-preserving qubit channel is either completely decohering or unital. For any given $\rho$, $\sigma\in\mS(d)$, if
\[
[\mE(\rho),\mE(\sigma)]=[\rho,\sigma]\neq0,
\]
the completely decohering case is excluded. Hence $\mE$ is unital. Note that the unital qubit channel $\mE$ acts on Bloch vectors can be rewritten as
\[
\bm r\longmapsto D\bm r
\]
for some $D=\operatorname{diag}(\lambda_1,\lambda_2,\lambda_3)$ with $|\lambda_i|\le1$. For
\[
\rho=\frac12(I+\bm r\cdot\bm\sigma),
\quad
\sigma=\frac12(I+\bm s\cdot\bm\sigma),
\]
we have
\[
[\rho,\sigma]
=
\frac{{\rm i}}{2}(\bm r\times\bm s)\cdot\bm\sigma.
\]
The identity
\[
(D\bm r)\times(D\bm s)
=
\operatorname{Cof}(D)\bm n
\]
shows that equality of the commutator norms requires
\[
|\operatorname{Cof}(D)\bm n|
=
|\bm n|,
\]
where $\operatorname{Cof}(D)=\operatorname{diag}(\lambda_2\lambda_3, \lambda_1\lambda_3, \lambda_1\lambda_2)$
and $|\lambda_i|\le1$. This implies that any nonzero component of $\bm n$ forces the corresponding product $|\lambda_i\lambda_j|$ to equal one. Hence two of the $|\lambda_i|$'s are equal to one. The complete-positivity conditions for a unital qubit channel then force the remaining canonical parameter also to have unit modulus. Therefore the Bloch transformation is a rotation, or equivalently, $\mE$ is a unitary channel.

\textit{Case 2: $d\ge3$.} In such a case, $\mE$ is either completely decohering or isotropic. The former is again excluded when we consider $\vec{\rho}=\{\rho, \sigma\}$ with $[\rho,\sigma]\neq0$. Thus $\mE$ is isotropic. For the unitary-type isotropic channel
\[
\mE(X)
=
tUXU^\dagger
+
(1-t)\frac{\tr(X)}{d}I,
\]
one has
\[
[\mE(\rho),\mE(\sigma)]
=
t^2U[\rho,\sigma]U^\dagger,
\]
and therefore
\[
C_p(\mE(\rho),\mE(\sigma))
=
t^2 C_p(\rho,\sigma).
\]
Since $C_p(\rho,\sigma)>0$, equality requires $|t|=1$. Complete positivity
imposes
\[
-\frac{1}{d^2-1}\le t\le1,
\]
so $t=1$, that is, $\mE$ is unitary.

For the transpose-type isotropic channel
\[
\mE(X)
=
tUX^TU^\dagger
+
(1-t)\frac{\tr(X)}{d}I,
\]
the same calculation gives
\[
C_p(\mE(\rho),\mE(\sigma))
=
|t|^2C_p(\rho,\sigma).
\]
This equality would again require $|t|=1$, whereas complete positivity requires
\[
-\frac{1}{d-1}
\le t\le
\frac{1}{d+1},
\]
which excludes $|t|=1$ for $d\ge3$. Namely, this type of isotropic channel cannot occur under the assumption~\eqref{freezing}.
We thus prove that \eqref{freezing} holds if and only if $\mE$ is a unitary operation for $d\ge2$, and that the same conclusion follows for $\mC'_p$ by an analogous argument.
\end{proof}


\section{Approximate Broadcasting and set coherence}\label{sec7}


In this section, we discuss the operational meaning of the set-coherence measure. Nonzero set coherence rules out perfect broadcasting, which motivates the approximate version of the problem: how closely can the marginals of a common approximate broadcasting channel reproduce the input state~\cite{ApproxBroadcasting}?  We then quantify, in terms of the SCMs, how the amount of set coherence constrains the optimal performance of approximate broadcasting. For simplicity, we restrict to a pair of pure qubit states.

Following Ref.~\cite{ApproxBroadcasting},  we
adopt the quality of
approximate broadcasting
defined there; for two
equally likely input qubit states $\vec{\rho}=\{\rho_1,\rho_2\}$, we define the optimal average local broadcasting fidelity (OALBF) as
\bea\label{g2}
g_2(\vec{\rho})=\sup_{\mE}\frac{1}{4}\sum_{i=1}^{2}\sum_{j=1}^{2}\tr\left(\rho_i\,\varrho_{ij}\right),
\eea
where the supremum is taken over all quantum channels mapping the input qubit to a two-qubit output system and
\[
\varrho_i=\mE(\rho_i),~\varrho_{i1}=\tr_B\varrho_i, ~\varrho_{i2}=\tr_A\varrho_i.
\]
Namely, $g_2$ is the largest average squared local fidelity between the input states and the two output marginals achievable by a common broadcasting channel. We define the corresponding broadcasting deficit as
\be
\epsilon_{\rm br}(\vec{\rho})=1-g_2(\vec{\rho}).
\ee
We consider
\bea\label{example}
|\psi_1\rangle=|0\rangle, ~|\psi_2\rangle=\cos\alpha\,|0\rangle+\sin\alpha\,|1\rangle,~0\leqslant\alpha\leqslant\frac{\pi}{2},~~~~~
\eea
and let $\rho_i=|\psi_i\rangle\langle\psi_i|,i=1,2$. Their squared overlap is $q=|\langle\psi_1|\psi_2\rangle|^2=\cos^2\alpha$.
For the total Bargmann gap $\mG_B(\vec{\rho})=\frac12\|[\rho_1,\rho_2]\|_2^2$,
a direct calculation gives $\mG_B=\frac14\sin^2(2\alpha)$,
or equivalently
\be
\mG_B=q(1-q).
\label{eq:CB-q}
\ee
For these two pure states, exact broadcasting is possible only at the two endpoints. When $\alpha=0$, the two input states are identical, whereas for $\alpha=\pi/2$ they are orthogonal. In both cases, $\mG_B=0$, $g_2=1$, $\epsilon_{\rm br}=0$. For $0<\alpha<\pi/2$, the two states are distinct and nonorthogonal, so perfect broadcasting is impossible when $\mG_B>0$, $g_2<1$.

\begin{figure}[ht]
    \centering
    (a)\\
    \includegraphics[width=0.85\linewidth]{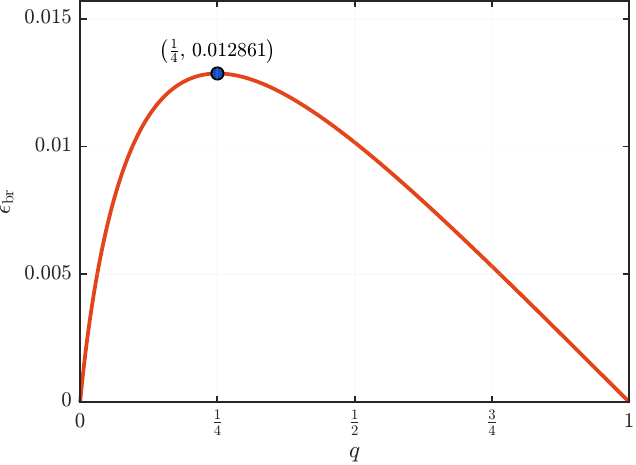}\\
    \vspace{3mm}(b)\\
    \includegraphics[width=0.85\linewidth]{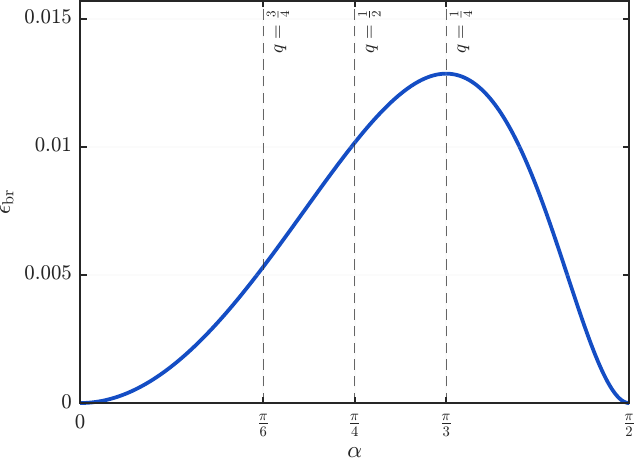}\\
    \caption{(a) The corresponding broadcasting deficit $\epsbr=1-g_2$ as a function of the squared overlap $q=|\la\psi_1|\psi_2\ra|^2$. $\epsbr$ reaches a maximum at $q=\frac14$. (b) Broadcasting deficit \(\epsilon_{\mathrm{br}}\) as a function of the angle \(\alpha\), with the special overlaps \(q=\frac{3}{4}\), \(q=\frac{1}{2}\), and \(q=\frac{1}{4}\) marked.}
    \label{fig:q}
\end{figure}

For two pure qubit states, the OALBF defined in Eq.~\eqref{g2} coincides with the optimal average single-copy fidelity of state-dependent quantum cloning studied in Ref.~\cite{Bruss1998}; the proof of this equivalence is given in Appendix~\ref{optimal}. Exploiting the optimal cloning fidelity derived there, we obtain the following closed-form expression for the OALBF (see Appendix~\ref{c} for the derivation):
\begin{widetext}
\bea\label{g2-2}
g_2(\vec{\rho})=\frac{1}{2}+\frac{\sqrt{2}}{32\sqrt{q}}(1+\sqrt{q})\left( 3-3\sqrt{q}+\sqrt{1-2\sqrt{q}+9q}\right) \sqrt{-1+2\sqrt{q}+3q+(1-\sqrt{q})\sqrt{1-2\sqrt{q}+9q}}.\label{eq:Fstar}
\eea
\end{widetext}
The function above can alternatively be denoted by $g_2(q)$, with the endpoint values understood by continuity as $g_2(0)=g_2(1)=1$. The corresponding broadcasting deficit is thus $\epsilon_{\rm br}(q)=1-g_2(q)$. For clarity, we plot the broadcasting deficit $\epsilon_{\rm br}(q)$ in Fig.~\ref{fig:q} (a); the case where $\epsilon_{\rm br}$ is viewed as a function of $\alpha$ (the angle between $|\psi_1\ra$ and $|\psi_2\ra$) is shown in Fig.~\ref{fig:q} (b).

\begin{figure}[htp]
    \centering
    \includegraphics[width=0.85\linewidth]{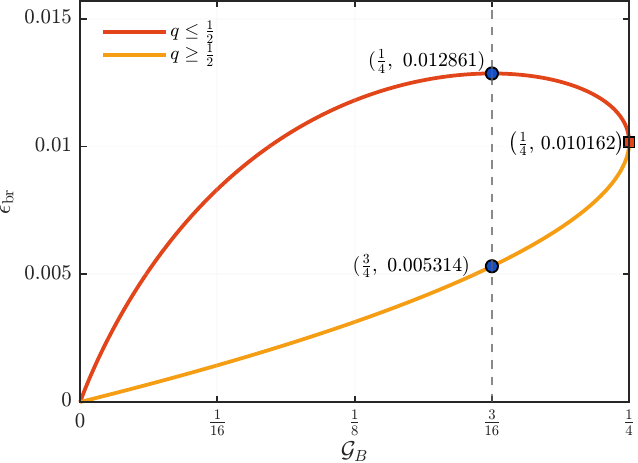}
    \caption{
    		The relation between the broadcasting deficit and the total Bargmann gap for $\vec{\rho}$ in Eq.~\eqref{example}.
    		The two branches correspond to squared overlaps $q\leqslant 1/2$ and $q\geqslant 1/2$, respectively. The two blue dots $(3/16, 0.012861)$ and $(3/16, 0.005314)$
    		highlight $\epsilon_{\mathrm{br},\mp}$ with the same $\mathcal G_B=3/16$, corresponding to $q=1/4$ and $q=3/4$, respectively. The broadcasting deficit attains its maximum at $q=1/4$, whereas the two branches meet at $\mathcal G_B=1/4$, corresponding to $q=1/2$ and $\epsilon_{\rm br}\approx 0.010162$. The interval $(0.010162, 0.012861)$ of $\epsilon_{\mathrm{br},-}$ corresponds to $\alpha\in(\pi/4, \pi/3)$, i.e., $q\in(1/4, 1/2)$.
    }
    \label{fig:validation}
\end{figure}

The overlap can now be eliminated in favor of the SCM. From Eq.~\eqref{eq:CB-q}, one has
$q_{\pm}=\frac{1\pm\sqrt{1-4\mathcal{G}_B}}{2}, 0\le \mathcal{G}_B\le \frac14,$
where \(q_-<\frac12\) and \(q_+>\frac12\). Accordingly, with $g_2$ defined as in Eq.~\eqref{g2-2}, the broadcasting deficit can be written as a piecewise function of the commutator-based SCM \(\mathcal{G}_B\):
\be
\begin{cases}
    \epsilon_{\mathrm{br},-}(\vec{\rho})=1-g_2\left(\dfrac{1-\sqrt{1-4\mathcal{G}_B(\vec{\rho})}}{2}\right), & q_-\leqslant\dfrac12,\\[3mm]
    \epsilon_{\mathrm{br},+}(\vec{\rho})=1-g_2\left(\dfrac{1+\sqrt{1-4\mathcal{G}_B(\vec{\rho})}}{2}\right), & q_+\geqslant\dfrac12.
\end{cases}
\label{eq:epsbranches}
\ee
For clarity, we plot the corresponding two analytic branches of the broadcasting deficit in Fig.~\ref{fig:validation}.

Fig.~\ref{fig:validation} reveals that the two state pairs corresponding to $q_-$ and $q_+$ for a given $\mG_B$ share the same
$\mathcal{G}_B$ but generally yield different broadcasting deficits; hence,
$\mathcal{G}_B$ alone does not uniquely determine the approximate
broadcasting performance. As shown in Appendix~\ref{c}, the upper branch $\epsilon_{\mathrm{br},-}$ is strictly increasing for $0\leqslant\mathcal G_B\leqslant 3/16$ and strictly decreasing for $3/16\leqslant\mathcal G_B\leqslant 1/4$, attaining its unique maximum at $\mathcal G_B=3/16$. In contrast, the lower branch $\epsilon_{\mathrm{br},+}$ is strictly increasing throughout $0\leqslant\mathcal G_B\leqslant 1/4$.

In other words, within these two
ranges of $q$, stronger set coherence implies a larger broadcasting deficit, as one
may expect. In contrast, for $q\in(\frac14,\frac12)$---equivalently, for
the upper branch $\epsilon_{\mathrm{br},-}$ with
$\frac{3}{16}\leqslant\mG_B\leqslant\frac14$---the broadcasting deficit
decreases as the set coherence increases. Thus, the amount of set coherence
reflects the broadcasting deficit within certain ranges of $q$.
At $\mathcal{G}_B=0$, corresponding to $q=0$ and $q=1$, the two
branches behave as
\bea
\epsilon_{\mathrm{br},\pm}(\mG_B)
=a_\pm\mG_B+o(\mG_B),
\eea
respectively, where $a_+=3/8-\sqrt{2}/4, a_-=1/4$.

We now consider the behavior of the broadcasting deficit associated with the robustness of set coherence
$C'_R$ and $C''_R$.
For the two pure qubit states in Eq.~\eqref{example}, as shown
in Appendix~\ref{app:robustness}, these measures take the analytic forms
\beax
\mC'_R(\vec{\rho})=\sqrt{q(1-q)},~~  \mC''_R(\vec{\rho})=\sqrt{\min\{q,1-q\}},
\eeax
respectively. Clearly, $0\le \mC'_R\le 1/2$, $0\le \mC''_R\le 1/\sqrt{2}$. Moreover, they satisfy
\beax
\mathcal G_B(\vec{\rho})=[\mC'_R(\vec{\rho})]^2=       [\mC''_R(\vec{\rho})]^2
\left[1-(\mC''_R(\vec{\rho}))^2\right] .
\eeax
It follows that
\bea
\epsilon_{\mathrm{br},\pm}(\vec{\rho})&=& 1-g_2\left(\frac{1\pm\sqrt{1-4[\mC'_R(\vec{\rho})]^2}}{2}\right),\\
\epsilon_{\mathrm{br},-}(\vec{\rho})
&=&1-g_2\left[ (\mC''_R(\vec{\rho}))^2\right],\\
\epsilon_{\mathrm{br},+}(\vec{\rho})
&=&1-g_2\left[1-(\mC''_R(\vec{\rho}))^2\right].
\eea

\begin{figure}[htp]
    \centering
 	(a)\\
 \includegraphics[width=0.85\linewidth]{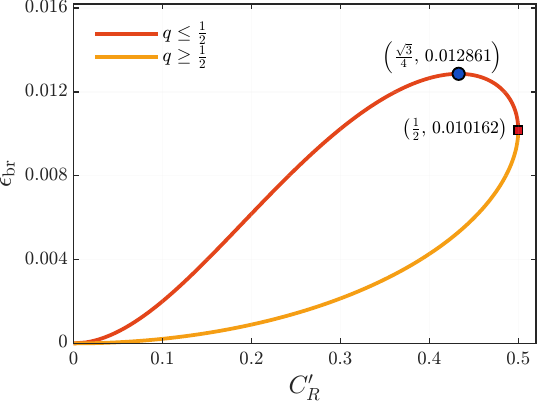}\\
 \vspace{3mm}(b)\\
 \includegraphics[width=0.85\linewidth]{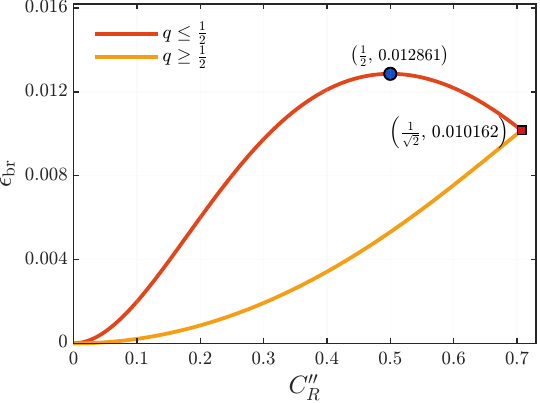}\\
    \caption{
        Broadcasting deficit $\epsilon_{\rm br}$
        as a function of
        (a) the mean robustness $\mC'_R$ and
        (b) the max robustness $\mC''_R$.
        The two analytic branches $\epsilon_{{\rm br},-}$ and $\epsilon_{{\rm br},+}$ correspond to
        $q\leqslant 1/2$ and $q\geqslant 1/2$, respectively.
        The marked blue dots indicate the maximum
        broadcasting deficit at $q=1/4$.
    }
    \label{fig:robustness_deficit}
\end{figure}

The resulting relations are illustrated in Fig.~\ref{fig:robustness_deficit}. Both robustness-based coherence measures exhibit the same two-branch structure as the commutator-based SCM $\mG_B$. Interestingly, all three SCMs attain their maximum at $q=1/2$, whereas the largest broadcasting deficit occurs at $q=1/4$; moreover, the monotonic intervals coincide with those of $\mathcal G_B$ in Fig.~\ref{fig:validation}.


\section{Conclusion and discussion}\label{sec8}


In this work, in the same spirit as the SIM investigated in~\cite{Guo2026pra}, we have established an axiomatic framework for set-coherence measures by the commutativity-preserving quantum channels which coincide with the free operations of set coherence. This axiomatic framework is the counterpart of the set imaginarity measure proposed in~\cite{Guo2026pra}, and these two frameworks are completely identical in form.

We have shown that the averaged robustness of set coherence  $\mC_R'$ is an SCM but is neither unified nor complete, whereas the maximal robustness $\mC_R''$ is a unified SCM but is neither complete nor strongly superadditive. We therefore have proposed an improved robustness of set coherence, $\mC'''_R$, which we show to be a complete SCM. Within the framework of state-set resource theories, these SCMs are closely analogous to the three types of robustness of set imaginarity introduced in~\cite{Guo2026pra}. We then presented two classes of complete set-coherence measures based on the Schatten $p$-norm of the commutator. The total Bargmann gap introduced in~\cite{Wang2026Bargmann} is indeed an SCM based on the Hilbert--Schmidt norm of the commutator. Moreover, all these norm-based SCMs are strongly superadditive.
For clarity, we compare all these SCMs in Table~\ref{tab:SCM-comparison}.
\begin{table}[htp]
    \caption{\label{tab:SCM-comparison}Comparison of the SCMs.} 
    \begin{ruledtabular}
    \begin{tabular}{cccc}
        SCM & Unified & Complete & Strongly superadditive\\ \colrule
        $\mC_p, \mC_p', \mG_B$
        &  \checkmark & \checkmark & \checkmark\\
        $\mC_R'$
        &  $\times$ & $\times$ & $\times$\\
        $\mC_R''$
        &  \checkmark & $\times$ & $\times$\\
        $\mC'''_R$
        &  \checkmark & \checkmark & $\times$
    \end{tabular}
\end{ruledtabular}
\end{table}

We further explored the operational significance of set coherence in the context of approximate quantum broadcasting. For pairs of pure qubit states, we established explicit relations between the broadcasting deficit and three SCMs: the commutator-based measure and two robustness-based measures. The analysis reveals that, although set coherence decreases while the approximate broadcasting deficit increases whenever the angle between the two states lies in $(\pi/4, \pi/3)$, the amount of set coherence quantified by our measures faithfully captures the approximate broadcasting deficit, and vice versa, whenever the two states are close to being either orthogonal or identical. We expect these results to stimulate further research on quantum resource theories and to advance our understanding of the role that resources play in set of states.



\begin{acknowledgements}
This work is supported by the National Natural Science Foundation of China under Grant Nos.~12471434 and 11971277, the Program for Young Talents of Science and Technology in Universities of Inner Mongolia Autonomous Region under Grant No. NJYT25010, the High-Level Talent Research Start-up Fund of Inner Mongolia University under Grant No. 10000-A260015/ 501, and the Inner Mongolia Autonomous Region Science and Technology Plan Projects under Grant Nos. 2025KYPT0098 and 2026KYPT0006.
\end{acknowledgements}


\appendix


\section{Proof of Lemma~\ref{monotonicity}}\label{a}


We distinguish between the cases $\dim\mH\geqslant3$ and $\dim\mH=2$.

\textit{Case 1}. $\dim\mH\geqslant3$. In such a case, a commutativity-preserving channel is either completely decohering or a nontrivial isotropic channel~\cite{Guo2013jpa-qd}. If $\mE$ is completely decohering, then its range is commutative, and hence $[\mE(\rho),\mE(\sigma)]=0$. The desired inequality is therefore immediate.

Otherwise, $\mE$ is an isotropic channel of the form
$\mE(A)=t\Gamma(A)+(1-t)\tr(A)I/d$,
where $\Gamma(A)=UAU^\dagger$ or
$\Gamma(A)=UA^T U^\dagger$ for some unitary $U$.
Since the identity part commutes with every operator,
$[\mE(\rho),\mE(\sigma)]
=t^2[\Gamma(\rho),\Gamma(\sigma)]$.
Moreover, $[\Gamma(\rho),\Gamma(\sigma)]$
equals $U[\rho,\sigma]U^\dagger$ in the former case
and $-U[\rho,\sigma]^T U^\dagger$ in the latter.
By the unitary and transpose invariance of
Schatten $p$-norms, we obtain
\begin{equation}
    \bigl\|[\mE(\rho),\mE(\sigma)]\bigr\|_p
    =|t|^2\bigl\|[\rho,\sigma]\bigr\|_p
    \leqslant\bigl\|[\rho,\sigma]\bigr\|_p,
\end{equation}
where the inequality follows from $|t|\leqslant1$.

\textit{Case 2}. $\dim\mH=2$.
If $\mE$ is completely decohering, the result is
immediate. Otherwise, $\mE$ is unital and acts on
Bloch vectors as $\br\mapsto T\br$, with
$\|T\|\leqslant1$.

For two qubit states
$\rho=(I+\br\cdot\boldsymbol{\sigma})/2$
and $\sigma=(I+\bs\cdot\boldsymbol{\sigma})/2$,
where $\boldsymbol{\sigma}$ denotes the vector
of Pauli operators, the Pauli identity
$(\br\cdot\boldsymbol{\sigma})
(\bs\cdot\boldsymbol{\sigma})
=(\br\cdot\bs)I+
{\rm i}(\br\times\bs)\cdot\boldsymbol{\sigma}$
gives
$[\rho,\sigma]
=\frac{{\rm i}}2(\br\times\bs)
\cdot\boldsymbol{\sigma}$.
Since
$[(\br\times\bs)\cdot\boldsymbol{\sigma}]^2
=|\br\times\bs|^2I$,
the two singular values of $[\rho,\sigma]$
are both $|\br\times\bs|/2$. Hence,
\begin{equation}
    \|[\rho,\sigma]\|_p
    =c_p|\br\times\bs|,
    \qquad
    c_p=2^{1/p-1},
    \label{eq:qubit_comm_norm}
\end{equation}
where $1\leqslant p\leqslant\infty$, with the convention
$c_\infty=1/2$.
Similarly,
$\|[\mE(\rho),\mE(\sigma)]\|_p
=c_p|T\br\times T\bs|$.

Let $s_1\geqslant s_2\geqslant s_3\geqslant0$ denote
the singular values of $T$.
Since $s_1=\|T\|\leqslant1$, the induced map
on the second exterior power satisfies
$\|\wedge^2T\|=s_1s_2\leqslant1$.
Consequently,
$|T\br\times T\bs|
=\|(\wedge^2T)(\br\wedge\bs)\|
\leqslant\|\wedge^2T\|\,
\|\br\wedge\bs\|
\leqslant|\br\times\bs|$.
Therefore,
\begin{equation}
    \|[\mE(\rho),\mE(\sigma)]\|_p
    \leqslant
    \|[\rho,\sigma]\|_p.
    \label{eq:qubit_monotonicity}
\end{equation}
This completes the proof.


\section{Optimal average local broadcasting fidelity for two qubit states}\label{optimal}


In this appendix, we calculate the OALBF for the two pure qubit states defined in Eq.~\eqref{example} and show that the resulting formulation agrees with the optimal state-dependent cloning fidelity reported in Ref.~\cite{Bruss1998}. We first give the following lemma.

\begin{lemma}
For the states $\vec{\rho}$ in~Eq. \eqref{example},

\bea\label{eq:result}
g_2(\vec{\rho})
&=&\max_{0\leqslant\phi\leqslant\pi/8}
\Biggl\{
\frac12+\frac{1+\sqrt q}{2}
\Biggl[
(1-\sqrt q)\cos 2\phi \nonumber\\
&&\qquad\qquad\qquad
+\frac{\sqrt q}{2}\sin 4\phi
\Biggr]
\Biggr\},
\eea
where $q=|\langle\psi_1|\psi_2\rangle|^2=\cos^2\alpha$.
\end{lemma}

\begin{proof}
Since the OALBF depends only on the overlap parameter \(q\), it is convenient to write \(g_2(\vec{\rho})\) simply as \(g_2(q)\). We prove Eq.~\eqref{eq:result} in four steps. First, we formulate the broadcasting optimization as a semidefinite program (SDP) and derive the corresponding dual problem. Second, we construct a family of feasible broadcasting channels, which yields a lower bound on the optimal fidelity. Third, we construct a candidate dual operator whose objective value matches this lower bound. Finally, we verify dual feasibility of this operator by exploiting the symmetries of the problem. Since the two bounds coincide, weak duality establishes Eq.~\eqref{eq:result}.

{Step 1: SDP formulation.} We first transform the input states into a symmetric form and reduce the broadcasting optimization to symmetric broadcasting channels. We then use the Choi representation to formulate the broadcasting optimization as an SDP and derive its dual problem.

For the set \(\vec{\rho}\) defined in Eq.~\eqref{example}, namely,
\[
|\psi_1\rangle=|0\rangle,\qquad
|\psi_2\rangle=\cos\alpha\,|0\rangle+\sin\alpha\,|1\rangle,
\quad 0\leqslant\alpha\leqslant\frac{\pi}{2},
\]
its OALBF is invariant under a common unitary transformation of the input states, provided the output system are transformed accordingly. We therefore apply the unitary operator
\[
U_\alpha=
\begin{pmatrix}
    \cos(\alpha/2)&\sin(\alpha/2)\\
    \sin(\alpha/2)&-\cos(\alpha/2)
\end{pmatrix},
\]
which transforms the input states into
\[
|\widetilde\psi_{1,2}\rangle
=\cos\frac{\alpha}{2}|0\rangle
\pm\sin\frac{\alpha}{2}|1\rangle.
\]
Writing \(\widetilde\rho_i=|\widetilde\psi_i\rangle\langle\widetilde\psi_i|\) and \(Z=\operatorname{diag}(1,-1)\), we obtain
\[
\widetilde\rho_2=Z\widetilde\rho_1 Z.
\]

Following the symmetric-broadcasting reduction established in Ref.~\cite{ApproxBroadcasting}, the optimization in Eq.~\eqref{g2} can be restricted, without loss of generality, to symmetric broadcasting channels whose outputs are invariant under the exchange of the two output systems. Since the two output marginals are identical for such channels, Eq.~\eqref{g2} reduces to
\bea
g_2(\vec\rho)
=\sup_{\mE~\text{is~symmetry}}
\frac12\sum_{i=1}^{2}\tr(\rho_i\rho_{i1}).
\eea

Since the auxiliary state $\tau$ in the
broadcasting process is fixed, we may
absorb its preparation into the channel
and define an effective broadcasting map
$\Lambda(\rho)=\mE(\rho\otimes\tau)$.
Thus, $\Lambda$ maps a single input qubit
to the two output systems $A$ and $B$.
Conversely, any such CPTP map can be
realized within the original broadcasting
formulation by discarding the initial
auxiliary system and applying $\Lambda$.
Therefore, optimizing over $\mE$
is equivalent to optimizing over all
effective channels $\Lambda$.

Let $J$ denote the unnormalized Choi matrix
of the broadcasting channel $\Lambda$,
acting on the Hilbert space
$\mH^A\otimes\mH^B
\otimes\mH^{\mathrm{in}}$,
where $A$ and $B$ denote the two output
systems and the superscript ``$\mathrm{in}$'' denotes the
input reference system. The complete positivity and trace preserving conditions are equivalent to $J\geqslant 0$ and $\tr_{AB}J=I_2$, respectively. For each input state $\widetilde\rho_i$, the channel output is represented by
$$
\mE(\widetilde\rho_i)
=\tr_{\mathrm{in}}
\left[J(I_{AB}\otimes\widetilde\rho_i^T)\right].
$$
Consequently, the local fidelity of the first output marginal can be written as
$$
\begin{aligned}
    \tr(\widetilde\rho_i\varrho_{i1})
    &=\tr\left[(\widetilde\rho_i\otimes I_B)\mE(\widetilde\rho_i)\right]=\tr\left[J(\widetilde\rho_i\otimes I_B
    \otimes\widetilde\rho_i^T)\right].
\end{aligned}
$$
Similarly, the fidelity of the second output marginal is
$$
\tr(\widetilde\rho_i\varrho_{i2})
=\tr\left[
J(I_A\otimes\widetilde\rho_i
\otimes\widetilde\rho_i^T)\right].
$$
Substituting these expressions into Eq.~\eqref{g2}, we obtain the primal SDP
\be
\begin{aligned}
    g_2(q)=\max_J\quad&
    \tr(K_\alpha J)\\
    \mathrm{s.t.}\quad&
    J\geqslant0,\qquad
    \tr_{AB}J=I_2,
\end{aligned}
\ee
where $K_\alpha=\frac14\sum_{i=1}^{2}\left(\widetilde\rho_i\otimes I_B+I_A\otimes\widetilde\rho_i\right)\otimes\widetilde\rho_i^T.$

To derive the dual SDP,
we introduce a Hermitian Lagrange multiplier
$Y$ associated with the trace-preserving
constraint $\tr_{AB}J=I_2$.
The corresponding Lagrangian is
\[
\begin{aligned}
    L(J,Y)
    &=\tr(K_\alpha J)
    +\tr
    \left[Y(I_2-\tr_{AB}J)\right]\\
    &=\tr Y+
    \tr
    \left[(K_\alpha-I_{AB}\otimes Y)J\right].
\end{aligned}
\]
Taking the supremum over $J\geqslant0$, we find that
the Lagrangian is bounded above if and
only if $K_\alpha-I_{AB}\otimes Y\leqslant 0$.
In this case, the supremum equals
$\tr Y$.
Therefore, the dual problem is
\[
\begin{aligned}
    \min_{Y=Y^\dagger}\quad&
    \tr Y\\
    \mathrm{s.t.}\quad&
    I_{AB}\otimes Y-K_\alpha\geqslant0.
\end{aligned}
\]

Every dual feasible operator $Y$
provides an upper bound on OALBF. Indeed, for any
primal feasible $J$ and dual feasible $Y$,
\[
\begin{aligned}
    \tr(K_\alpha J)
    &\leqslant\tr[(I_{AB}\otimes Y)J]\\
    &=\tr[Y\tr_{AB}J]\\
    &=\tr Y.
\end{aligned}
\]
Thus $g_2(q)\leqslant\tr Y$.
To establish the claimed global optimality,
it suffices to construct a feasible
broadcasting channel and a dual feasible
operator attaining the same objective value.

{Step 2: Construction of a feasible broadcasting channel.} We construct a feasible symmetric broadcasting
channel to obtain a lower bound on $g_2(q)$.
Since the two input states have been brought into
a symmetric form, we consider an isometry
$V_t:\mathbb C^2\to\mathbb C^2\otimes\mathbb C^2$
whose outputs lie in the symmetric subspace,
defined by
\[
\begin{aligned}
    V_t|0\rangle
    &=\cos t\,|00\rangle+\sin t\,|11\rangle,\\
    V_t|1\rangle
    &=\frac{|01\rangle+|10\rangle}{\sqrt{2}},
\end{aligned}
\]
where $0\leqslant t\leqslant\pi/4$.

The two output vectors are orthonormal, so
$V_t^\dagger V_t=I_2$.
Consequently,
$\Lambda_t(\rho)=V_t\rho V_t^\dagger$
defines a CPTP channel.
Furthermore, if $W_{AB}$ denotes the
swap operator on the two output qubits,
then $W_{AB}V_t=V_t$.
It follows that
$W_{AB}\mE_t(\rho)W_{AB}^\dagger
=\mE_t(\rho)$,
and hence the two output marginals coincide.

For convenience, let $S=\sqrt q=\cos\alpha$,
$a=\cos t$, and $c=\sin t$.
For the transformed input states
$|\widetilde\psi_\pm\rangle
=u|0\rangle\pm v|1\rangle$,
where $u=\cos(\alpha/2)$ and
$v=\sin(\alpha/2)$, the corresponding
joint outputs are
\[
|\Phi_\pm\rangle
=u(a|00\rangle+c|11\rangle)
\pm\frac{v}{\sqrt2}(|01\rangle+|10\rangle).
\]
Tracing out either output qubit gives
the identical marginal states
\[
\varrho_{\pm,A}=
\begin{pmatrix}
    u^2a^2+v^2/2 &
    \pm uv(a+c)/\sqrt2\\
    \pm uv(a+c)/\sqrt2 &
    u^2c^2+v^2/2
\end{pmatrix}.
\]

For each input state, the local fidelity is
$F_\pm=\langle\widetilde\psi_\pm|
\varrho_{\pm,A}|\widetilde\psi_\pm\rangle$.
Since the two output marginals coincide,
and the two input states yield the same
local fidelity $F_+=F_-$, the average
local fidelity achieved by the specific
broadcasting channel $\Lambda_t$ is
$F(t)=(F_++F_-)/2=F_\pm$.
Direct calculation gives
\bex
F(t)=u^4a^2+u^2v^2c^2
+\frac{v^2}{2}
+\sqrt2\,u^2v^2(a+c).
\eex
Using $u^2=(1+S)/2$ and
$v^2=(1-S)/2$, we obtain
\bex
\begin{aligned}
    F(t)=&
    \frac{(1+S)^2}{4}a^2+\frac{1-S^2}{4}c^2\\
    &+\frac{1-S}{4}+\frac{\sqrt2(1-S^2)}4(a+c).
\end{aligned}
\eex
Since $\Lambda_t$ is a valid CPTP broadcasting channel, it follows from Eq.~\eqref{g2} that $F(t)\leqslant g_2(q)$. For subsequent calculations, we introduce $A=(1+S)^2/4$, $C=(1-S^2)/4$, $D=(1-S)/4$, and $h=\sqrt2(1-S^2)/8$. Then the fidelity takes the compact form
\[
F(t)=Aa^2+Cc^2+D+2h(a+c).
\]
Under the change of variables $2\phi=\pi/4-t$, this becomes
\be
F(t)=\frac12+\frac{1+S}{2} \left[(1-S)\cos 2\phi+\frac S2\sin 4\phi\right].
\ee
This coincides with the cloning fidelity on the right-hand side of Eq.~\eqref{eq:result}. Since $\mE_t$ is a valid
broadcasting channel, we obtain
\be
g_2(q)\geqslant
\max_{0\leqslant t\leqslant\pi/4}F(t).
\ee
Equivalently, using $2\phi=\pi/4-t$, this lower bound is precisely the maximum on the right-hand side of Eq.~\eqref{eq:result}. It remains to establish the matching upper bound.

{Step 3: Construction of a dual certificate.} We now construct a candidate dual operator and verify
its feasibility. For $0<S<1$, the fidelity $F(t)$
is strictly concave on $0\leqslant t\leqslant\pi/4$,
since $F''(t)<0$ throughout this interval.
Moreover, $F'(0)>0$ and $F'(\pi/4)<0$,
so there exists a unique maximizer
$t_*\in(0,\pi/4)$.
Differentiating $F(t)$ and setting its derivative
to zero gives
\bex
\sqrt{2}S\sin 2t_*
=(1-S)(\cos t_*-\sin t_*).
\eex
For subsequent calculations, we write
$a=\cos t_*$ and $c=\sin t_*$.
Equivalently, the stationarity condition
takes the form
\be\label{eqcondition}
(A-C)ac=h(a-c)
\ee

Motivated by this relation, we construct
a Hermitian operator on the input qubit,
\bex\label{eq:dual_candidate}
    Y_*=
    \begin{pmatrix}
        y_0&0\\
        0&y_1
    \end{pmatrix},
\eex
where $y_0=A+h/a=C+h/c$ and
$y_1=D+h(a+c)$.
The equality between the two expressions
for $y_0$ follows from
Eq.~\eqref{eqcondition}. Using $a^2+c^2=1$ and the stationarity condition in Eq.~\eqref{eqcondition}, we have $y_0=Aa^2+Cc^2+h(a+c)$. Consequently,
$$
Y_*=y_0+y_1=Aa^2+Cc^2+D+2h(a+c)=F(t_*).
$$
Thus, $Y_*$ has the same objective value as the optimal fidelity within the
cloning family.
To establish global optimality,
it remains to verify the dual feasibility
condition
\be\label{eq:dual_feasibility}
    I_{AB}\otimes Y_*-K_\alpha\geqslant0.
\ee

{Step 4: Dual feasibility and optimality.} To verify the dual feasibility condition, we exploit two symmetries of the operator $T=I_{AB}\otimes Y_*-K_\alpha$.
First, the two output systems enter $K_\alpha$ symmetrically, so exchanging $A$ and $B$ leaves $K_\alpha$ unchanged.
Since $I_{AB}\otimes Y_*$ is also invariant under this exchange, we have $[T,W_{AB}\otimes I_{\mathrm{in}}]=0$,
where $W_{AB}$ denotes the swap operator on the two output systems.

Second, the transformed input states satisfy
$\widetilde{\rho}_-=Z\widetilde{\rho}_+Z$.
Let $P=Z\otimes Z\otimes Z$. Since conjugation
by $P$ interchanges the two terms in the sum
defining $K_\alpha$, we have
\[
P K_\alpha P=K_\alpha.
\]
Moreover, $Y_*$ is diagonal and hence commutes
with $Z$, which implies
$P(I_{AB}\otimes Y_*)P=I_{AB}\otimes Y_*$.
Consequently,
\[
PTP=T,
\]
or equivalently, $[T,P]=0$.

The two symmetry operators commute with each
other and have eigenvalues $\pm1$. Since $T$
commutes with both operators, it preserves
their joint eigenspaces and hence admits a
block-diagonal representation in a common
eigenbasis. The eight-dimensional Hilbert
space therefore decomposes into four invariant
subspaces of dimensions $3,3,1,1$.

To identify these subspaces explicitly, we adopt the
tensor-factor ordering $\mH^A\otimes
\mH^B\otimes\mH^\mathrm{in}$
and introduce the symmetric and antisymmetric Bell
states $|\Psi^\pm\rangle=(|01\rangle\pm|10\rangle)/\sqrt{2}$.
The two three-dimensional joint eigenspaces are
\[
\begin{aligned}
    \mathcal H_+
    &=\operatorname{span}\{
    |000\rangle,|110\rangle,
    |\Psi^+\rangle|1\rangle\},\\
    \mathcal H_-
    &=\operatorname{span}\{
    |001\rangle,|\Psi^+\rangle|0\rangle,
    |111\rangle\},
\end{aligned}
\]
with joint swap-parity eigenvalues $(+1,+1)$
and $(+1,-1)$, respectively.
The remaining one-dimensional subspaces are
spanned by $|\Psi^-\rangle|0\rangle$ and
$|\Psi^-\rangle|1\rangle$, respectively, with
joint eigenvalues $(-1,-1)$ and $(-1,+1)$. In this symmetry-adapted basis, $T$ is unitarily
equivalent to
\[
T\simeq M_+\oplus M_-\oplus\lambda_0\oplus\lambda_1.
\]
In the ordered bases of $\mH_+$ and
$\mH_-$ specified above, the two
$3\times3$ blocks take the forms
\[
M_+=
\begin{pmatrix}
    y_0-A & 0 & -h\\
    0 & y_0-C & -h\\
    -h & -h & y_1-D
\end{pmatrix}
\]
and
\[
M_-=
\begin{pmatrix}
    y_1-C & -h & 0\\
    -h & y_0-\frac{1+S}{4} & -h\\
    0 & -h & y_1-\frac{(1-S)^2}{4}
\end{pmatrix}.
\]
The remaining two scalar blocks are $\lambda_0=y_0-(1+S)/4$ and $\lambda_1=y_1-(1-S)/4$.

Since unitary transformations preserve positive semidefiniteness, the dual feasibility condition $T\geqslant0$ is equivalent to requiring $M_+\geqslant0$, $M_-\geqslant0$, and $\lambda_0,\lambda_1\geqslant0$. We now verify these conditions using the stationary relation satisfied by the optimal parameter $t_*$.

We begin with $M_+$. Recall that the stationarity condition gives $y_0=A+h/a=C+h/c$, while $y_1=D+h(a+c)$. Substituting these relations yields
\[
M_+=h
\begin{pmatrix}
    1/a & 0 & -1\\
    0 & 1/c & -1\\
    -1 & -1 & a+c
\end{pmatrix}.
\]

For any real vector $(u,v,w)^T$, its quadratic form satisfies
\[
\begin{aligned}
    &(u,v,w)M_+(u,v,w)^T\\
    &\quad~~~~~~~=h\left[
    \frac{(u-aw)^2}{a}
    +\frac{(v-cw)^2}{c}
    \right]\geqslant0.
\end{aligned}
\]
Since $a,c,h>0$, this proves $M_+\geqslant0$.

We next consider $M_-$. For simplicity, denote its diagonal entries by $U=y_1-C$, $H=y_0-(1+S)/4$, and $W=y_1-(1-S)^2/4$.
Then
\[
M_-=
\begin{pmatrix}
    U&-h&0\\
    -h&H&-h\\
    0&-h&W
\end{pmatrix}.
\]
Using the definitions of $A,C,D,h$ and the candidate dual operator $Y_*$,
we find
\[
\begin{aligned}
    U&=\frac{1-S}{8}
    \left[\sqrt2(1+S)(a+c)-2S\right],\\
    H&=\frac{S(1+S)}4+\frac ha,\\
    W&=\frac{S(1-S)}4+h(a+c).
\end{aligned}
\]
For $0<S<1$, we have $H,W>0$. Moreover, since $a+c\geqslant1$ and $\sqrt2(1+S)>2S$, we also have $U>0$.

Since $U,W>0$, applying the Schur complement
with respect to the diagonal block
$\operatorname{diag}(U,W)$ gives
\be\label{eq:schur}
M_-\geqslant0
~\Longleftrightarrow~
H\geqslant h^2\left(\frac1U+\frac1W\right).
\ee

To verify this inequality, let $r=c/a\in(0,1)$. The stationarity condition implies $(1-S)(1-r)\sqrt{1+r^2}=2\sqrt2 Sr$.
Since $U,W>0$, Eq.~\eqref{eq:schur} is equivalent to $\Delta:=UWH-h^2(U+W)\geqslant0$. Using the stationarity relation, we obtain
\[
\begin{aligned}
    \Delta={}&
    \frac{S^3(1+S)(1+r)}
    {64(1-r)(1+r^2)^2}\\
    &\times[(1+r)^2-S(1-r)^2]\\
    &\times[S(1+r)^2-(1-r)^2].
\end{aligned}
\]

The prefactor and the first bracket are strictly
positive. To verify the remaining condition
$S(1+r)^2\geqslant(1-r)^2$, we use the stationarity
relation to obtain
\[
S=\frac{(1-r)\sqrt{1+r^2}}
{(1-r)\sqrt{1+r^2}+2\sqrt{2}r}.
\]
Substituting this expression, we find that the required
inequality reduces to
$\sqrt{2}\sqrt{1+r^2}\geqslant1-r$,
which holds for $0<r<1$.
Hence, $\Delta\geqslant0$ and consequently
$M_-\geqslant0$.

Finally, the two scalar constraints hold since $y_0-(1+S)/4=H>0$ and $y_1-(1-S)/4=h(a+c)>0$. Together with $M_+\geqslant0$ and $M_-\geqslant0$, this establishes Eq.~\eqref{eq:dual_feasibility}, so $Y_*$ is dual feasible.

Since $\Lambda_{t_*}$ is a valid broadcasting channel and $\tr Y_*=F(t_*)$, weak duality gives
\begin{equation}
    F(t_*)\leqslant g_2(q)
    \leqslant\tr Y_*=F(t_*).
\end{equation}
Therefore, $g_2(q)=F(t_*)=\max_{0\leqslant t\leqslant\pi/4}F(t) $. Introducing $\phi=\pi/8-t/2\in[0,\pi/8]$, we recover Eq.~\eqref{eq:result} for $0<q<1$. At $q=0$, the two input states are orthogonal and can be cloned perfectly, whereas at $q=1$ they coincide and can also be broadcast perfectly. In both cases, $g_2(q)=1$, consistent with Eq.~\eqref{eq:result}. This completes the proof.
\end{proof}

In order to obtain Eq.~\eqref{g2-2}, we let $S=|\langle\psi_1|\psi_2\rangle|=\sqrt{q}$ and regard $g_2$ as a function of $S$. Then, the corresponding fidelity before optimization is given by
\beax
g_2(S,\phi)
=~~~~~~~~~\hspace{5cm}\nonumber\\
\frac12+\frac12(1+S)
\left[
(1-S)\cos 2\phi
+\frac{S}{2}\sin 4\phi
\right],~~~~
\eeax
where $0\leqslant\phi\leqslant\pi/8$. The OALBF is then $g_2(S)=\max_{\phi}g_2(S,\phi)$. We write $x=\sin2\phi$, then $\cos2\phi=\sqrt{1-x^2}$. Since $0\leqslant\phi\leqslant\pi/8$, we obtain
\bea\label{eq:g2Sx}
g_2(S,x)
=
\frac12+\frac12(1+S)\sqrt{1-x^2}
(1-S+Sx).~~~
\eea
The stationarity condition $\partial g_2/\partial x=0$ gives $2Sx^2+(1-S)x-S=0$, whose unique positive root is
\begin{equation}
	x_*(S)=\frac{S-1+\sqrt{1-2S+9S^2}}{4S}.\label{eq:xstar}
\end{equation}
Substituting $x_*(S)$ into Eq.~\eqref{eq:g2Sx}, we obtain
\be
\begin{aligned}
	g_2^{\mathrm{opt}}(S)
	&=g_2\bigl(S,x_*(S)\bigr)\\
	&=\frac{1}{2}+\frac{1+S}{2}
	\sqrt{1-x_*^2(S)}\bigl(1-S+Sx_*(S)\bigr),
\end{aligned}
\ee
which, upon setting $S=\sqrt q$ and simplifying,
yields Eq.~\eqref{g2-2}.


\section{Monotonicity of the two branches $\epsilon_{\mathrm{br},\pm}$}\label{c}


To determine the monotonicity of $\epsilon_{\mathrm{br},\pm}$, we apply the envelope theorem. Since the optimal parameter $x_*(S)$ satisfies the stationarity condition
$\partial g_2/\partial x=0$, the chain rule gives
\[
g_2'(S)=
\left.
\frac{\partial g_2(S,x)}{\partial S}
\right|_{x=x_*(S)} =\frac12\sqrt{1-x_*^2}[(1+2S)x_*-2S].
\]
Writing $D=\sqrt{1-2S+9S^2}$ and using Eq.~\eqref{eq:xstar}, we obtain $(1+2S)x_*-2S=[(1+2S)D-(6S^2+S+1)]/(4S)$. Since both terms in the numerator are positive, their difference has the same sign as
\beax
(1+2S)^2D^2-(6S^2+S+1)^2=8S^2(2S-1).
\label{eq:factorization}
\eeax
Therefore,
$\operatorname{sgn}g_2'(S)
=\operatorname{sgn}(2S-1)$
for $0<S<1$. Hence, $g_2(S)$ is strictly decreasing for $0<S<1/2$ and strictly increasing for $1/2<S<1$, attaining its unique minimum at $S=1/2$, or equivalently $q=1/4$.

Since the broadcasting deficit is given by $\epsilon_{\mathrm{br}}(q)=1-g_2(q)$, its monotonicity is opposite to that of the OALBF. Recall that $\mathcal G_B=q(1-q)$ is invariant under $q\mapsto1-q$, giving rise to two branches corresponding to $q\leqslant1/2$ and $q\geqslant1/2$. Moreover,
\be
    \frac{d\mathcal G_B}{dq}=1-2q,
\ee
so $\mathcal G_B$ is strictly increasing for $0<q<1/2$ and strictly decreasing for $1/2<q<1$.

Using $q=S^2$ and the chain rule, we obtain
\be
    \frac{d\epsilon_{\mathrm{br}}}
    {d\mathcal G_B}
    =
    -\frac{g_2'(\sqrt q)}
    {2\sqrt q(1-2q)}.
\ee
Here $g_2'(S)$ denotes the derivative with respect to $S$. For $1/2<q<1$, we have $g_2'(\sqrt q)>0$ and $1-2q<0$, implying $d\epsilon_{\mathrm{br}}/d\mathcal G_B>0$. Thus, the $q>1/2$ branch is strictly increasing with $\mathcal G_B$. For $0<q<1/2$, the denominator is positive, while $g_2'(\sqrt q)$ changes sign at $q=1/4$. Consequently, the broadcasting deficit increases with $\mathcal G_B$ for $0<q<1/4$ and decreases for $1/4<q<1/2$, attaining its unique maximum at $\mathcal G_B=(1/4)(3/4)=3/16$.


\section{Robustness of set coherence for qubit states}\label{app:robustness}

Let $\vec{\rho}=\{\rho_1,\rho_2\}$ be a set of two pure qubit states, where $\rho_j=|\psi_j\rangle\langle\psi_j|$ and $q=\tr(\rho_1\rho_2)$. In the Bloch representation, $\rho_j=\frac12(I+\mathbf r_j\cdot\boldsymbol{\sigma})$ with $|\mathbf r_j|=1$. For a reference basis specified by a unit vector $\mathbf n$ as in Eq.~\eqref{robustness}, the robustness of coherence is~\cite{Napoli2016}
\bex
C_R^{(\mathbf n)}(\rho_j)=\sqrt{1-(\mathbf r_j\cdot\mathbf n)^2}.
\eex
Since $C_R^{(\mathbf n)}(\rho_j)$ depends only on the squared inner product $(\mathbf r_j\cdot\mathbf n)^2$, the optimization is invariant under $\mathbf r_j\mapsto-\mathbf r_j$ and $\mathbf n\mapsto-\mathbf n$. We therefore introduce $\theta\in[0,\pi/2]$ through $\cos\theta=|\mathbf r_1\cdot\mathbf r_2|=|2q-1|$. Consequently, $\sin\theta=2\sqrt{q(1-q)}$.

For the mean robustness, let $\theta_j\in[0,\pi/2]$ be defined by $\cos\theta_j=|\mathbf r_j\cdot\mathbf n|$. Then $C_R^{(n)}(\rho_j)=\sin\theta_j$. The triangle inequality for angular distance between unoriented directions gives $\theta\leqslant\theta_1+\theta_2$. We now show that
\[
\sin\theta\leqslant\sin\theta_1+\sin\theta_2.
\]
Indeed, if $\theta_1+\theta_2\leqslant\pi/2$, the inequality follows from the monotonicity of the sine function and the addition formula. If $\theta_1+\theta_2\geqslant\pi/2$, then $\sin\theta_1+\sin\theta_2\geqslant1 \geqslant\sin\theta$. Equality holds when $\mathbf n$ is parallel to either Bloch axis (i.e., $\mathbf n$ parallel to the line through $\mathbf r_1$ or $\mathbf r_2$). Consequently,
\begin{equation}
    C_R'(\vec{\rho})
    =\frac12\min_{|\mathbf n|=1}
    \sum_{j=1}^{2}\sin\theta_j
    =\frac12\sin\theta
    =\sqrt{q(1-q)}.
\end{equation}

For the max robustness, the angular triangle inequality implies
\beax
    \theta\leqslant\theta_1+\theta_2
    \leqslant2\max\{\theta_1,\theta_2\}.
\eeax
Since $\sin\theta_j$ is increasing
on $[0,\pi/2]$, we obtain $\max\{\sin\theta_1,\sin\theta_2\}\geqslant\sin\frac{\theta}{2}$. The lower bound is attained by choosing $\mathbf n$ along the bisector of the smaller angle between the lines spanned
by $\mathbf r_1$ and $\mathbf r_2$. This gives $\theta_1=\theta_2=\theta/2$. Therefore,
\begin{equation}
    \begin{aligned}
        C_R''(\vec{\rho})
        &=\min_{|\mathbf n|=1}
        \max_{j=1,2}\sin\theta_j\\
        &=\sin\frac{\theta}{2}=\sqrt{\frac{1-|2q-1|}{2}}\\
        &=\sqrt{\min\{q,1-q\}}.
    \end{aligned}
\end{equation}



\end{document}